\documentclass[sigconf]{acmart}

\usepackage{hyperref}
\usepackage{balance}       %
\usepackage{graphicx}      %
\usepackage{color}
\usepackage{booktabs}
\usepackage{textcomp}
\usepackage{subcaption}
\usepackage{enumerate}
\usepackage{xcolor}
\usepackage{lipsum}%
\usepackage{makecell}
\usepackage{multicol}
\usepackage{multirow}
\usepackage{array}
\usepackage{verbatimbox}
\usepackage{enumitem}
\usepackage{amsmath}
\usepackage{stfloats}
\usepackage{graphicx}
\usepackage{amsthm}
\usepackage{listings}
\usepackage{caption} 
\usepackage{xspace}
\usepackage[linesnumbered]{algorithm2e}
\usepackage{algpseudocode}
\usepackage{tabularx}
\usepackage{arydshln}
\usepackage[bottom]{footmisc}
\usepackage{stackengine}
\usepackage{placeins}

\newcommand*{\eg}{\textit{e.g.},\xspace}
\newcommand*{\ie}{\textit{i.e.},\xspace}

\newcommand*{\etal}{\textit{et~al.}\xspace}

\newcolumntype{L}[1]{>{\raggedright\let\newline\\\arraybackslash\hspace{0pt}}m{#1}}
\newcolumntype{C}[1]{>{\centering\let\newline\\\arraybackslash\hspace{0pt}}m{#1}}
\newcolumntype{R}[1]{>{\raggedleft\let\newline\\\arraybackslash\hspace{0pt}}m{#1}}

\makeatletter
\def\thickhline{%
  \noalign{\ifnum0=`}\fi\hrule \@height \thickarrayrulewidth \futurelet
   \reserved@a\@xthickhline}
\def\@xthickhline{\ifx\reserved@a\thickhline
               \vskip\doublerulesep
               \vskip-\thickarrayrulewidth
             \fi
      \ifnum0=`{\fi}}
\makeatother

\makeatletter
\def\thickhlinespace{%
  \addlinespace[1ex]
  \noalign{\ifnum0=`}\fi\hrule \@height \thickarrayrulewidth \futurelet
   \reserved@a\@xthickhline
   \addlinespace[1ex]
   }
\def\@xthickhlinespace{\ifx\reserved@a\thickhline
               \vskip\doublerulesep
               \vskip-\thickarrayrulewidth
             \fi
      \ifnum0=`{\fi}}
\makeatother

\newlength{\thickarrayrulewidth}
\newlength\Origarrayrulewidth

\definecolor{downredcolor}{HTML}{e31a1c}
\definecolor{upgreencolor}{HTML}{33a02c}

\definecolor{DarkGreen}{HTML}{5DAC81}

\newcommand\projectname{MindfulAgents\xspace}

\copyrightyear{2026}
\acmYear{2026}
\setcopyright{cc}
\setcctype{by}
\acmConference[CHI '26]{Proceedings of the 2026 CHI Conference on Human Factors in Computing Systems}{April 13--17, 2026}{Barcelona, Spain}
\acmBooktitle{Proceedings of the 2026 CHI Conference on Human Factors in Computing Systems (CHI '26), April 13--17, 2026, Barcelona, Spain}
\acmPrice{}
\acmDOI{10.1145/3772318.3791817}
\acmISBN{979-8-4007-2278-3/2026/04}

\begin{document}

\title{\projectname: Personalizing Mindfulness Meditation via an Expert-Aligned Multi-Agent System}

\author{Mengyuan “Millie” Wu}
\affiliation{%
  \institution{Columbia University}
  \city{New York}
  \state{NY}
  \country{USA}
}
\email{mw3209@columbia.edu}

\author{Zhihan Jiang}
\authornote{These authors contributed equally as second authors.}
\affiliation{%
  \institution{Columbia University}
  \city{New York}
  \state{NY}
  \country{USA}
}
\email{zj2445@cumc.columbia.edu}

\author{Yuang Fan}
\authornotemark[1]
\affiliation{%
  \institution{Columbia University}
  \city{New York}
  \state{NY}
  \country{USA}
}
\email{yf2676@columbia.edu}

\author{Richard Feng}
\affiliation{%
  \institution{St. Margaret's Episcopal School}
  \city{San Juan Capistrano}
  \state{CA}
  \country{USA}
}
\email{richardqfeng@gmail.com}

\author{Sahiti Dharmavaram}
\affiliation{%
  \institution{Columbia University}
  \city{New York}
  \state{NY}
  \country{USA}
}
\email{sd3976@columbia.edu}

\author{Mathew Polowitz}
\affiliation{%
  \institution{Carnegie Mellon University}
  \city{Pittsburgh}
  \state{PA}
  \country{USA}
}
\email{mathew@equahealth.io}

\author{Shawn Fallon}
\affiliation{%
  \institution{Carnegie Mellon University}
  \city{Pittsburgh}
  \state{PA}
  \country{USA}
}
\email{shawn@equahealth.io}

\author{Bashima Islam}
\affiliation{%
  \institution{Worcester Polytechnic Institute}
  \city{Worcester}
  \state{MA}
  \country{USA}
}
\email{bislam@wpi.edu}

\author{Lizbeth Benson}
\affiliation{%
  \institution{Institute for Social Research, University of Michigan}
  \city{Ann Arbor}
  \state{MI}
  \country{USA}
}
\email{libenson@umich.edu}

\author{Irene Tung}
\affiliation{%
  \institution{California State University Dominguez Hills}
  \city{Carson}
  \state{CA}
  \country{USA}
}
\email{itungphan@csudh.edu}

\author{David Creswell}
\affiliation{%
  \institution{Carnegie Mellon University}
  \city{Pittsburgh}
  \state{PA}
  \country{USA}
}
\email{creswell@andrew.cmu.edu}

\author{Xuhai Xu}
\affiliation{%
  \institution{Columbia University}
  \city{New York}
  \state{NY}
  \country{USA}
}
\email{xx2489@cumc.columbia.edu}

\renewcommand{\shortauthors}{}
\renewcommand{\shorttitle}{}

\begin{abstract}
Mindfulness meditation is a widely accessible and evidence-based method for supporting mental health. Despite the proliferation of mindfulness meditation apps, sustaining user engagement remains a persistent challenge. 
Personalizing the meditation experience is a promising strategy to improve engagement, but it often requires costly and unscalable manual effort.
We present \projectname, a multi-agent system powered by large language models that:
(1) generates guided meditation scripts based on an expert-established mindfulness framework,
(2) encourages users' reflection on emotional states and mindfulness skills, and 
(3) enables real-time personalization of the mindfulness meditation experience for each user. In a {formative lab study} (N=13), \projectname significantly improved in-session engagement ($p = 0.011$) and self-awareness ($p=0.014$), as well as reduced momentary stress ($p = 0.020$). Furthermore, a {four-week deployment study} (N=62) demonstrated a notable increase ($p = 0.002$) in long-term engagement and level of mindfulness ($p = 0.023$). Participants reported that \projectname offered more relevant meditation sessions personalized to individual needs in various contexts, supporting sustained practice. Our findings highlight the potential of LLM-driven personalization for enhancing user engagement in digital mindfulness meditation interventions.
\end{abstract}


\maketitle

\section{Introduction}
\label{sec:introduction}

\begin{figure*}[t]
  \centering
  \includegraphics[width=\textwidth,keepaspectratio]{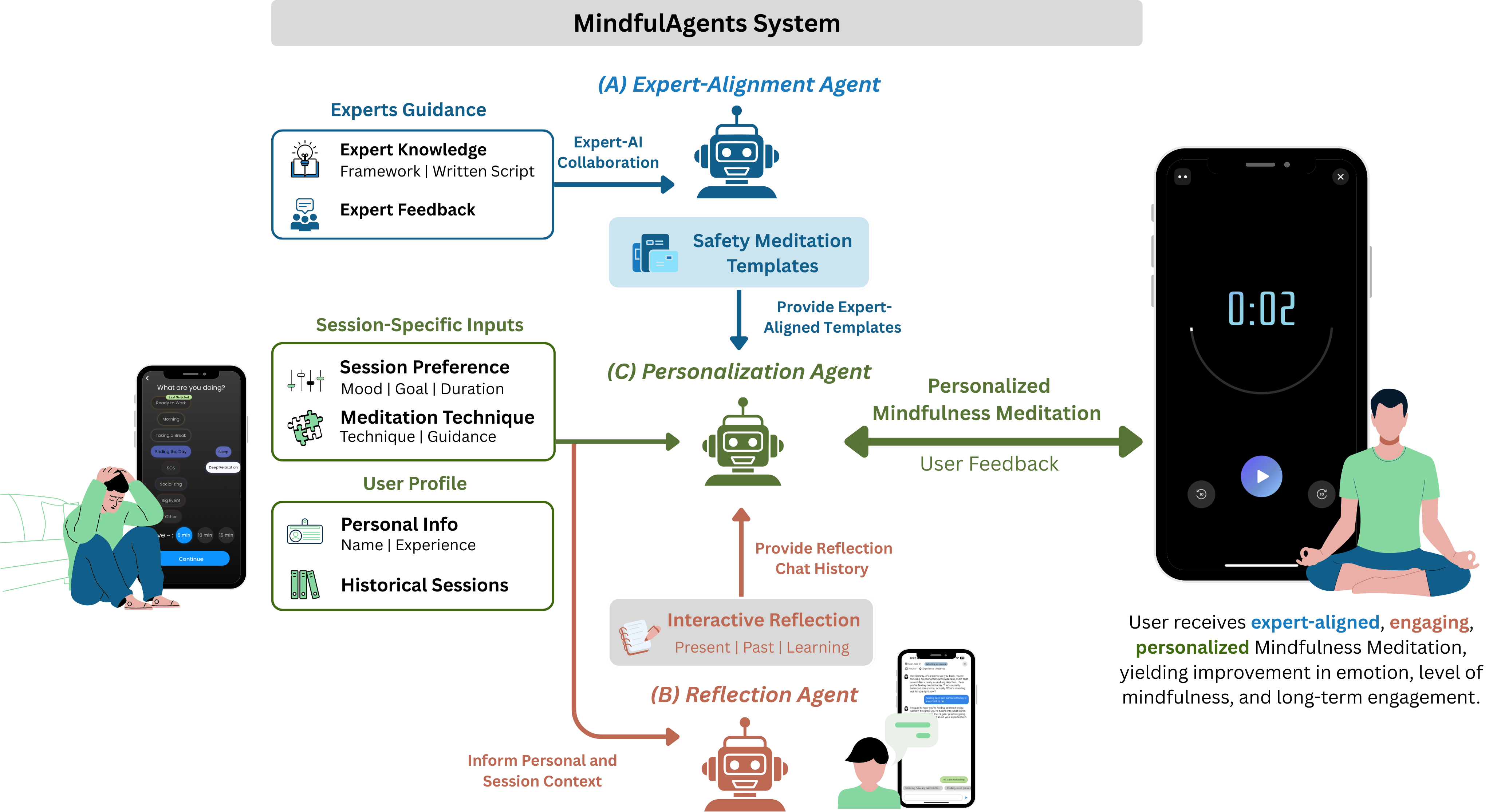}
  \caption{\projectname, a multi-agent system that provides expert-aligned, engaging, and personalized mindfulness meditation experience. The multi-agent system consists of:
  \textbf{(A)} an Expert-Alignment Agent that utilizes expert knowledge and feedback in the creation of safety meditation templates.
  \textbf{(B)} a Reflection Agent that encourages users to ponder their current emotional state, past meditation experiences, and technique-related questions, and
  \textbf{(C)} a Personalization Agent that builds on both the safety templates and reflective inputs to generate sessions that are user-specific and resonant.}
  \Description{Figure 1 illustrates the MindfulAgents system, a multi-agent framework for expert-aligned, engaging, and personalized mindfulness meditation. On the left, inputs include expert knowledge and feedback, session preferences (mood, goal, duration), meditation techniques, and user profile data (personal info, historical sessions). At the top, the Expert-Alignment Agent produces safety meditation templates through collaboration with experts. At the bottom, the Reflection Agent engages users in interactive reflection based on present, past, and learning contexts. At the center, the Personalization Agent integrates safety templates, user profile, and reflection history to generate personalized mindfulness meditations. On the right, users receive personalized sessions via a mobile app, with feedback loops informing further personalization.}
  \label{fig:teaserimage}
\end{figure*}

Mental health challenges are both widespread and costly, affecting more than one billion individuals across the world~\cite{WHO2025MentalHealth}. Recent reports show that at any given moment, one in eight people is living with a mental health disorder~\cite{WHO2025MentalHealth}.
Mindfulness meditation (referred to as ``meditation'' throughout the paper) is one of the most well-established, evidence-based approaches for improving mental health \cite{zhangMindfulnessbasedInterventionsOverall2021}.
It is the practice of present-centered awareness of thoughts, feelings, or sensations acknowledged and accepted free of judgment \cite{moraalvarezEffectsWebbasedMindfulness2023}. 
Extensive research has demonstrated that meditation can enhance emotional regulation, mitigate depression and anxiety symptoms, and improve overall well-being across diverse populations \cite{sanileviciMindfulnessBasedStressReduction2021, finlayImpactNeurologicalDisability2022, carlsonMindfulnessBasedInterventionsPhysical2012, goyal_meditation_2014}.
Moreover, meditations are broadly accessible, as they require minimal resources and do not involve strenuous physical activity, making them appropriate for individuals across a wide range of physical abilities and socioeconomic statuses~\cite{spears_perceptions_2017}. 
Guided meditation—delivered either in person by trained instructors or through digital apps—has become one of the most prominent ways people learn and sustain meditation practice, offering structured sessions that provide direction, pacing, and support~\cite{creswell_meditation_2025, macrynikola_impact_2024}.

Despite high accessibility, the efficacy of meditation is ultimately determined by sustained engagement, which is critical for yielding robust and lasting outcomes~\cite{bowlesDoseresponseEffectsReported2025, tanakaImpactContinuedMindfulness2023, cearnsEffectsDosePractice2023}. 
However, current meditation practices face a significant engagement gap. In the United States, 92.0\% of people report awareness of meditation, yet only 15.1\% have ever practiced it. Among those who have tried, 70.7\% failed to establish a regular practice \cite{baumelObjectiveUserEngagement2019}. One study of 26 popular meditation apps found that approximately 95\% of active users discontinued within the first month \cite{baumelObjectiveUserEngagement2019}.
Low or inconsistent engagement in meditation practice has been associated with being less beneficial for depressive symptom reduction, resilience, and mood regulation \cite{cearnsEffectsDosePractice2023, tanakaImpactContinuedMindfulness2023}.
Human meditation trainers can foster greater engagement by offering personalized guidance and adapting to individual needs \cite{camacho_impact_2023, wildeboer_relationship_2016, winter_engagement_2022}, yet such in-person coaching protocols remain financially and logistically inaccessible. 

The recent rise of large language models (LLMs) brings a promising opportunity to democratize personalized meditation to individual needs due to the models' ability to generate dynamic content and adapt interactively \cite{Lawrence2024, ma2024understanding,li_systematic_2023,van_agteren_systematic_2021}.
For example, \citet{kumar_large_2024} developed LLM-based informational chatbots that improved adoption and in-session engagement during mindfulness tutorials. 
 \citet{nguyen_ai-driven_2024} showed that personalizing meditation using LLM could potentially improve the user experience. However, the current work mainly focused on a single meditation session and did not aim to address the long-term engagement challenge through personalization.

To bridge this gap, we conducted formative co-design sessions with mindfulness experts (N=4), a comprehensive structure for meditation guidance with well-defined concepts and procedures~\cite{chinMindfulnessInterventionsImprove2021, lindsayHowMindfulnessTraining2018}. From these sessions, experts identified three key factors to promote engagement and retention.
First, reliability and consistency are essential, as practitioners are more likely to sustain regular practice when guidance is dependable; for LLM-generated scripts, this means avoiding hallucinations and maintaining a stable format across sessions to foster trust and routine~\cite{nguyen_ai-driven_2024}.
Second, structured reflection before practice is critical, since experienced instructors often prompt practitioners to consider their progress, challenges, and intentions before meditating—a step shown to deepen learning and strengthen long-term adherence~\cite{manigaultExaminingPracticeEffects2021}. 
Third, personalization to user context is necessary to ensure that meditation guidance remains relevant and engaging over time; this involves adapting to key dimensions such as mood, goals, technique, session duration, preferred level of guidance, and practice history~\cite{hubertyEvaluationMoodCheckin2021, mashederAmGettingSomething2020, bowenEvaluatingTheoryDrivenMessaging2025}.

Guided by these insights, we developed \textbf{\projectname}, a system that adopts a multi-agent setup to achieve expert-aligned, reflective, and personalized meditation experience (see Figure~\ref{fig:teaserimage}). Our system integrates the following three tightly coupled components: 
(1) An \textit{Expert-Alignment Agent} that collaborates with meditation experts and provides expert-vetted safety templates to guide the meditation guidance generation, ensure adherence to the UM framework, and prevent harmful or misleading outputs.
(2) A \textit{Reflection Agent} that encourages users to ponder their current emotional state, past meditation experiences, and technique-related questions. This reflective step not only primes users for meditation practices but also provides meaningful context for personalization.
(3) A \textit{Personalization Agent} that builds on both the safety templates and reflective inputs from the other two agents to generate sessions that are user-specific and resonant. By adapting the expert-aligned template to real-time user contexts, the system ensures the meditation practice is both personally meaningful and structurally safe. We carefully curated the input elements to the system through an ablation study.

To evaluate the effectiveness of \projectname, we first conducted a within-subjects {formative} study (N=13), which showed that \projectname significantly increased in-session engagement ($p = 0.011$), deepened self-awareness ($p=0.014$), and reduced momentary stress ($p = 0.020$).
Building on these findings, we further conducted another 4-week between-subjects {deployment} study in the wild (N=62).
Our results revealed that \projectname significantly outperformed the baseline in long-term engagement that sustained over time ($p=0.002$). It also provides significantly higher benefits on the level of mindfulness ($p=0.006$).
Participants also reported being motivated to continue because \projectname provided reflective dialogue, was more closely tailored to their personal context, and offered more diverse meditation experiences.

Overall, our contributions are: 
\begin{itemize}
\item We built \projectname, a multi-agent system that provides expert-aligned, personalized mindfulness meditation guidance to provide a more accessible and engaging meditation experience.%
\item We comprehensively evaluated our system through an offline ablation study, a {formative lab study} (N=13), and a {4-week deployment study} (N=62) on the effectiveness of the multi-agent-powered mindfulness meditation interventions.

\item We discussed how our work illuminates future design directions such as emotion accompaniment through personalization, the synergy between reflection and mindfulness, as well as the balance between expert-aligned safety and LLMs' generative novelty.
\end{itemize}

\section{Background and Related Work}
\label{sec:related_work}

{We ground our work at the intersection of mindfulness-based interventions, user engagement in digital meditation, and the emerging use of large language models (LLMs) in mental health support. Below, we first summarize evidence on the mental health benefits of mindfulness meditation, then discuss the enduring challenge of sustaining engagement in digital mindfulness programs, and finally review recent progress and limitations in applying LLMs for mental health and mindfulness contexts.}

\subsection{Mindfulness Meditation for Mental Health and Well-Being}
\label{sub:related_work:mindfulness}

Mindfulness meditation has been widely recognized as an effective intervention for improving mental health. A substantial body of evidence confirms its role in reducing depression, anxiety, and especially stress~\cite{mak_enhancing_2015, flett_mobile_2019}. Structured Mindfulness-Based Interventions (MBI), such as Mindfulness-Based Stress Reduction (MBSR) and Mindfulness-Based Cognitive Therapy (MBCT), further contribute to enhancing overall well-being by improving self-compassion, sleep quality, and sense of accomplishment~\cite{galante_mindfulness-based_2021, black_mindfulness_2015, neff_pilot_2013}. The benefits are particularly pronounced for populations experiencing elevated symptoms, such as patients with depression, anxiety, or chronic pain~\cite{goyal_meditation_2014}, as well as individuals in high-stress environments such as schools and workplaces~\cite{vonderlin_mindfulness-based_2020, dunning_research_2019}. Notably, these positive outcomes can demonstrate long-term effects, persisting months after program completion~\cite{shankland_improving_2021}.

In recent years, the therapeutic benefits of mindfulness have largely extended to digital formats, which broaden accessibility through web and mobile platforms.
Randomized controlled trials investigating web-based mindfulness programs demonstrate positive mental health outcomes across diverse populations~\cite{sommers-spijkerman_new_2021, mak_enhancing_2015}, and mobile application-based interventions have been shown to reduce psychological distress and enhance resilience, proving particularly beneficial for student cohorts~\cite{schulte-frankenfeld_app-based_2022, Balci2023, flett_mobile_2019}. Furthermore, digital mindfulness interventions provided crucial support during periods of heightened stress and limited resources, such as the COVID-19 pandemic, by helping individuals manage rumination and mood ~\cite{wang_web-based_2023}.

\subsection{Engagement Challenge in Meditation}
\label{sec:related_work:engagement}
Despite the evidence-based benefits and the broad awareness of mindfulness meditation (92.0\% of the US population reports awareness of meditation, according to \cite{baumelObjectiveUserEngagement2019}), only 15.1\% have ever practiced it. Among those who have tried, establishing a consistent habit has been a challenge, with 70.7\% failing to maintain a regular practice \cite{baumelObjectiveUserEngagement2019}. The proliferation of digital tools has not resolved this issue; although over 2,500 meditation apps were available worldwide~\cite{LaRosa}, attrition rates remain severe. Studies consistently reveal sharp declines in user activity following initial onboarding periods~\cite{winter_engagement_2022, flett_mobile_2019}. For example, a study of 26 popular meditation apps found that approximately 95\% of active users discontinued practice within the first month \cite{baumelObjectiveUserEngagement2019}.

This engagement gap is critical because the psychological and behavioral benefits of meditation are strongly linked to sustained practice duration and frequency~\cite{bowlesDoseresponseEffectsReported2025, tanakaImpactContinuedMindfulness2023, cearnsEffectsDosePractice2023}. Dose–response analyses suggest that approximately 160 hours of lifetime mindfulness practice are required to achieve clinically significant improvements in psychological distress and life satisfaction, with around 270 hours needed for these gains to stabilize over time \cite{bowlesDoseresponseEffectsReported2025}. Empirical evidence aligns with this threshold: a cross-sectional study of participants who completed an 8-week meditation intervention found that only those who maintained long-term practice reported significantly lower depression and higher resilience compared to those who discontinued \cite{tanakaImpactContinuedMindfulness2023}. Similarly, longitudinal data from digital meditation platforms show that consistent practice, e.g., four to seven days per week over a 14-month period, yielded the largest and most durable improvements in mood, equanimity, and stress resilience \cite{cearnsEffectsDosePractice2023}. Together, these findings underscore that only through regular, long-term engagement can meditation yield robust and enduring outcomes.

{High attrition in digital health underscores that many aspects of engagement depend on personalized support, including meaningful contact, timely reminders, and assistance on overcoming early hurdles~\cite{eysenbachLawAttrition2005}.
Research on digital health interventions shows that users are more likely to engage and benefit when systems are personalized to their characteristics, states, and context \cite{kankanhalliUnderstandingPersonalizationHealth2021, mohrBehavioralInterventionTechnology2014}. 
In the context of meditation, personalization mechanisms show similar promise: mindfulness trainers can foster greater engagement by offering personalized support and adapting practices to individual needs, often outperforming unguided digital interventions \cite{sommers-spijkerman_new_2021}.} The inclusion of trainer-led components in web-based mindfulness programs has been shown to significantly increase adherence and retention compared to those that are only self-guided \cite{winter_engagement_2022}. Another study found that participants receiving personalized recommendations completed substantially more activities than those who received only general support \cite{camacho_impact_2023}. 
However, these engagement benefits come with a substantial cost. While research on mobile mental health interventions consistently shows that guided programs achieve higher uptake and adherence, the considerable time and resources required for training, compensation, and management render labor-intensive coaching models impractical for large-scale deployment, especially in low-resource settings~\cite{schlapferEngagementRelaxationMindfulness2024, renfrewInfluenceHumanSupport2021}.
This tension creates a critical need for interventions that offer personalized support without sacrificing scalability.

\subsection{LLM for Mental Health and Mindfulness Meditation}
\label{sub:related_work:llm_mh}
The recent surge of LLMs offers a transformative opportunity to democratize mental well-being support, especially in regions with limited mental health resources. With their ability to demonstrate human-like empathy, LLMs show potential to provide emotional support, mitigate the stigma associated with seeking traditional mental health services, and reduce healthcare costs \cite{Lawrence2024, Guo2024Large, ma2024understanding}.
The application of LLMs in mental health spans several areas from assessment to education and intervention. For assessment, they have shown significant promise in diagnosing psychiatric conditions and performing affective analysis  (
\eg \cite{xu_mental-llm_2024,tao_classifying_2023, verma_ai-enhanced_2023, stigall_large_2024, lossio-ventura_comparison_2024}).
For education, LLMs are being explored to improve the training experience for mental health professionals and students~\cite{barish_automatically_2023, sezgin_clinical_2023}, with a focus on mitigating hallucination and unreliability through expert-aligned materials~\cite{spallek_can_2023, lai_supporting_2024, hu_psycollm_2024}. 
As an intervention tool, recent research has started to employ LLM-empowered chatbots to deliver intervention (\eg~\cite{ma2024understanding,li_systematic_2023,he_physician_2024, heinz2025randomized}).
While many of these LLM-based mental health chatbots enhance personalization through prompt engineering and user data (\eg demographics, emotional states, psychiatric questionnaires) \cite{santos_therapist_2020} or specialized datasets like clinical records and social media data \cite{das_conversational_2022}, their reliability remains a significant concern. The inherent variability and potential for hallucinations in LLM-generated responses introduce critical safety risks. This can lead to inaccurate advice, posing potential health hazards if not carefully monitored and controlled \cite{heston_safety_2023,lawrence2024opportunities}.

While LLMs are increasingly applied to the broader mental health domain, their integration into structured mindfulness practice is relatively underexplored and has focused on education and awareness rather than personalized intervention.
Recent work utilized LLM–based informational and reflective chatbots with tutorial videos from web-based mindfulness programs, successfully promoting meditation awareness and adoption \cite{kumar_large_2024}. However, these systems typically provide static content and lack mechanisms for ongoing, personalized intervention. {For instance, commercial meditation apps largely position LLM integration as opaque, proprietary recommenders that route users to pre-recorded catalogs, rather than providing live adaptive meditation \cite{BalanceGuidedMeditation2025,AIPrinciplesHeadspace2025}.}
Some studies have begun to explore the potential of LLM-driven personalization in mindfulness support \cite{nguyen_ai-driven_2024}.
Closer to our work, Nguyen \etal fine-tuned LLaMA-2 using participant-specific context to generate customized sessions \cite{nguyen_ai-driven_2024}. While this work illustrates the promise of personalized mindfulness interventions, the study reported no significant improvement in user ratings compared to non-AI baselines, largely due to reliability challenges such as frequent hallucinations in the generated guidance. 
Meanwhile, most prior work has focused on a single meditation session, and the long-term engagement aspect has been underexplored. {Thus, while LLM-based personalization is promising for scaling engagement-critical functions, current approaches still fall short of delivering safe, reliable, expert-aligned personalized meditation.}

{
Taken together, prior efforts highlight some key gaps: 
First, sustaining user engagement in digital meditation interventions remains a challenge.
By matching goals, preferences, and moments of need, \emph{personalization} can improve engagement (see Sec.~\ref{sec:related_work:engagement}). Human-supported interventions can achieve this level of personalization, but such approaches do not scale.
At the same time, LLM-based personalization introduces \emph{reliability} risks, which require expert-aligned constraints to bound model behavior.
In this work, we pair these two elements: a multi-agent system that provides scalable meditation sessions for \emph{personalization} while enforcing expert alignment and guardrails for \emph{reliability}, comprehensively evaluated through both a formative study and an in-the-wild deployment.
}

\section{Formative Co-Design with Mindfulness Experts}
\label{sec:co-design}
To inform the development of \projectname and ensure its reliability, we collaborated with mindfulness experts to design the system together.
We first introduce the mindfulness framework identified by experts and adopted in our system, Unified Mindfulness (Sec.~\ref{sub:co-design:framework}), and then introduce the co-design process, together with the design guidelines learned through the process (Sec.~\ref{sub:co-design:co-design}).

\subsection{A Mindfulness Framework: Unified Mindfulness (UM)}
\label{sub:co-design:framework}

{Unified Mindfulness (UM) is a structured meditation framework that was initially developed by Shinzen Young, a well-known meditation coach and neuroscience research consultant~\cite{unifiedmindfulness}, and later enhanced and enriched by other meditation practitioners and researchers~\cite{chin2019mindfulness,manigaultExaminingPracticeEffects2021}.
The framework offers a structured 14-day training program designed to teach meditation concepts in a systematic and comprehensive way, with comprehensive validation.  {Unified Mindfulness has been evaluated in multiple randomized controlled trials by a leading mindfulness laboratory, showing reliable effects on neural, physiological, and behavioral markers of well-being \cite{lindsayMindfulnessTrainingReduces2019, lindsayHowMindfulnessTraining2018, slutskyMindfulnessTrainingImproves2019, chinMindfulnessTrainingReduces2019, manigaultExaminingPracticeEffects2021, chinMindfulnessInterventionsImprove2021a}.} These trials demonstrated that participants exhibited significantly reduced cortisol levels and blood pressure reactivity \cite{lindsayHowMindfulnessTraining2018, lindsayAcceptanceLowersStress2018}, as well as improved attentional control, focus, mood, and overall happiness in daily life \cite{slutskyMindfulnessTrainingImproves2019, lindsayHowMindfulnessTraining2018}.}

The UM framework grounds meditation practice on the three core attentional skills, including \textit{Concentration Power}, \textit{Sensory Clarity}, and \textit{Equanimity} \cite{unifiedmindfulness}.
The primary technique - \textit{See, Hear, Feel} - is defined as a practice where individuals deconstruct their moment-to-moment experience into these three sensory modalities: (1) \textit{See} for visual phenomena (external sights and internal images); (2) \textit{Hear} for auditory stimuli (external sounds and internal talk); and (3) \textit{Feel} for somatic sensations (physical and emotional feelings).
By continuously labeling experiences as they arise, practitioners develop the capacity to observe phenomena with internal clarity, non-reactivity, and equanimity. This methodology facilitates both formal, seated meditation and informal integration into daily activities.

\subsection{Co-Design Process and Design Guidelines}
\label{sub:co-design:co-design}

To address the persistent engagement challenges in digital meditation and mitigate LLM's safety risk, we first iteratively co-designed with 4 mindfulness experts (3 males, 1 female, all with 10+ years of mindfulness and meditation experience). All experts were familiar with the UM framework and agreed upon the adoption of UM throughout the design process. {It is noteworthy that \projectname's technical pipeline takes the framework-specific curriculum, expert demonstrations, and conversational materials as modular inputs. Technically, our pipeline is also compatible with materials from other expert-designed meditation frameworks (e.g., Mindfulness-based Stress Reduction (MBSR) \cite{kabat-zinnFullCatastropheLiving2005}, Mindfulness-based Cognitive Therapy (MBCT) \cite{segalMindfulnessbasedCognitiveTherapy2002}), while the underlying data generation, modeling, and evaluation procedures remain unchanged. In this work, we focus on UM as a first instantiation and system capability proof-of-concept, based on its strong empirical validation.} Over a five-month period, we engaged in an unstructured, iterative co-design process. This involved regular meetings, typically with a subset of one to three experts at a time, to facilitate deep discussion. After we introduced an early-stage prototype, in each following session, the experts would review the latest iteration of the system, propose new features, and help inspect potential usability, safety, or experiential issues. This cycle of feedback and refinement continued until we and the experts felt that the design had reasonably converged and addressed many of the core user needs and safety considerations we had identified.

\subsubsection{Design Guidelines}
We iteratively refined our system from a vanilla LLM to the final \projectname architecture. The detailed steps and corresponding illustration are provided in the Appendix (Figure~\ref{fig:design_iteration} and Table~\ref{tab:design-iterations}). Early iterations focused on expert alignment (finetuning on expert-generated scripts and adding format) and framework-alignment checks to reduce hallucinations and voice API errors. We then moved to a personalization-agent design for real-time generation with user inputs, adding safety guardrails to keep behavior reliable while reducing latency. Finally, we introduced a dedicated reflection agent to strengthen therapeutic alliance by drawing on user context, reflective conversations, past sessions, and related concept clarifications. 
Through the co-design process, we converged on a multi-agent architecture that better addressed the intertwined needs of engagement, reliability, and personalized support. We distilled three design guidelines (DGs) that guided system development aimed at enhancing engagement and reliability in mindfulness practice.

\textbf{DG1: Ensuring Consistency in Meditation Guidance Content Generation.}

{We began our design iteration process with a minimal prototype: a generic LLM-based meditation coach (Iteration~1 in Figure~\ref{fig:design_iteration} Table~\ref{tab:design-iterations}). Mindfulness experts noticed that the agent often used vague wellness language and did not reliably reflect the underlying mindfulness framework. To build trust and maintain reliability, they emphasized that meditation guidance should closely follow the UM curriculum, minimize hallucinations, and preserve a consistent pedagogical progression. This feedback motivated us to introduce multiple safety layers (Iterations~2–4), adopting expert-aligned material finetuning and format checks rather than purely free-form generation. To ensure reliability and expert-alignment, we grounded our system in multiple resources from real-world UM practices: }

(1) \textit{Concept introduction.} UM framework provides detailed explanations of its core concepts, accompanied by definition materials~\cite{unifiedmindfulness}. For instance, Appendix Tab.~\ref{tab:um_techniques} lists out core UM mindfulness techniques\cite{UMWiki}. These resources form a knowledge base intended to support more conceptually accurate content and reduce risks of hallucination or misrepresentation.
(2) \textit{Meditation scripts.} We incorporated standard meditation lesson scripts from a 14-day training program in a flagship meditation app \footnote{We used a commercially available Unified Mindfulness training app \cite{EquaHealth2025} as a source of standard lesson scripts.}, which introduces how to practice core UM techniques in a structured, teaching-oriented format.
In addition, specific goal-oriented practice scripts (\eg handling stress, building connection, and staying present, see the complete list of goals in Appendix Tab.~\ref{tab:category_goals}), which experts identified as vital for demonstrating how UM can assist users with daily life challenges. While the 14-day training program ensures systematic exposure to expert-aligned concepts, the goal-oriented scripts provide practical relevance but require significant personalization and adaptation by the LLM.
(3) \textit{Instructor–Practitioner interactions.} Experts also provided transcripts of conversations from in-person meditation sessions that capture how expert instructors responded to students' questions. These examples provided valuable examples for the LLM system on how to shape the authentic and supportive instructor–learner dynamics.

\textbf{DG2: Encourage Reflection Prior to Meditation Practice.}

{Experts emphasized that effective mindfulness teaching requires adapting examples and metaphors to the individual and building a sense of relational safety. By Iteration~6, the generated content was perceived as technically correct but emotionally flat. \textit{``The content is aligned with the existing framework, but it feels like the same script no matter who I am. A real teacher remembers what we talked about last time and connects today’s practice to that.''} (E3, UM instructor) In real-world mindfulness teaching practice, human instructors often begin sessions by prompting practitioners to reflect on their current experiences, past progress, and conceptual understanding—an approach shown to deepen learning and sustain long-term engagement. Incorporating a structured reflection step into our system can provide similar benefits, helping users connect meditation practice to their broader life journey. In later iterations, this took the form of a Reflection Agent (Iterations~7–8) that engages users in a brief, structured check-in and connects the upcoming practice to prior sessions and life context.}

Expert feedback highlighted three dimensions of reflection that are particularly valuable: (1) \textit{Current state of being}, where users articulate their present circumstances and emotions, fostering resonance and a sense of companionship with the system; (2) \textit{Past meditation journey}, where users review their prior practice, successes, and challenges, promoting continuity and self-awareness; and (3) \textit{UM concept review}, where users clarify or revisit key terms and ideas, reducing confusion and reinforcing conceptual learning.  
Embedding such reflection can not only strengthen user engagement but also generate meaningful feedback loops for personalization, enabling our system to adapt its guidance over time in ways that are both contextually sensitive and experientially grounded.  

\textbf{DG3: Capturing Core Personalization Inputs.}  
Drawing from our co-design sessions, the experts emphasized that effective personalization hinges on capturing user input across several key dimensions:
(1) \textit{Mood} tracking allows the system to adapt sessions to users' affective state, which can strengthen engagement and foster a sense of intimacy \cite{hubertyEvaluationMoodCheckin2021}.
(2) Connecting meditation to user-defined \textit{goals} and daily life provides motivation and increases the likelihood of sustained practice \cite{mashederAmGettingSomething2020}.
(3) Supporting multiple meditation \textit{techniques} ensures scientific grounding while allowing users to explore proper techniques suited to their preferences.
(4) Flexible session \textit{duration} is essential, as difficulty fitting practice into a busy daily schedule is one of the primary reasons for attrition \cite{bowenEvaluatingTheoryDrivenMessaging2025, manigaultExaminingPracticeEffects2021}.
(5) The \textit{level of guidance detail} must adapt to practitioners' expertise: novice practitioners often prefer more structured guidance, while experienced meditators benefit from greater autonomy.
(6) A user's \textit{prior practice history} ensures continuity by building on past sessions, avoiding repetition, and scaffolding their learning journey over time.
{Together, these inputs create a foundation for tailoring mindfulness experiences to individual needs. At the same time, clinician stakeholders further emphasized safety, scope of practice, and avoiding over-promising therapeutic benefits. They noted that the agent should not attempt to resolve acute crises on its own, but instead maintain clear boundaries and hand off to human or crisis resources when risk emerges. \textit{``The agent shouldn’t try to `fix' panic attacks on its own. It needs clear boundaries and a way to hand off to human or crisis resources when risk shows up.''} (E2, psychologist) Their feedback drove the addition of explicit safety checks, escalation responses, and conservative language around mental health claims, implemented through safety resources, red-flag detection prompts, and fallback responses integrated into the Expert-Alignment and Personalization agents (Iterations~5–7).}

\section{\projectname: An Expert-Aligned Multi-Agent System for Personalized Meditation}
\label{sec:system}

Following the three core DGs, we develop \projectname, a multi-agent system consisting of an \textit{Expert-Alignment Agent}, a \textit{Reflection Agent}, and a \textit{Personalization Agent}.
As shown in Fig.~\ref{fig:teaserimage}, the pipeline begins with the \textit{Expert-Alignment Agent} (Sec.~\ref{sub:system:expert_aligned_agent}), which generates safe meditation guidance templates grounded in expert knowledge and feedback.
When initiating a session, users start with a brief dialogue with the \textit{Reflection Agent} (Sec.~\ref{sub:system:reflective_agent}) to capture their present state, recall past sessions, and clarify terminology as needed.
The \textit{Personalization Agent} (Sec.~\ref{sub:system:personalization_agent}) then tailors the safety meditation templates from \textit{Expert-Alignment Agent} by synthesizing the user's current input, historical meditation data, and reflective chat history.
The finalized meditation script is converted into natural-sounding speech and delivered back to the user in real-time.
The \projectname interface (Sec.~\ref{sub:system:interface}) provides a smooth user flow that goes through these steps.

Overall, we used the OpenAI GPT-4.1 model family to develop the three agents~\cite{openai41}. All session information, including user inputs, reflective exchanges, and personalized outputs, is stored in a Pinecone vector database~\cite{Pinecone} to enable efficient retrieval for personalization, evaluation, and future system adaptation.
We employed the ElevenLabs platform~\cite{ElevenLabs} for the auto audio generation.

\subsection{Expert-Alignment Agent: Safety Meditation Template Creation}
\label{sub:system:expert_aligned_agent}
To ensure LLM's consistent generation that is aligned with the mindfulness framework (\textbf{DG1}), safety meditation templates are developed through a three-phase expert-AI collaborative process to ensure high-quality, safe, and effective mindfulness guidance for improved engagement. 

\begin{figure*}
    \centering
    \includegraphics[width=0.85\linewidth]{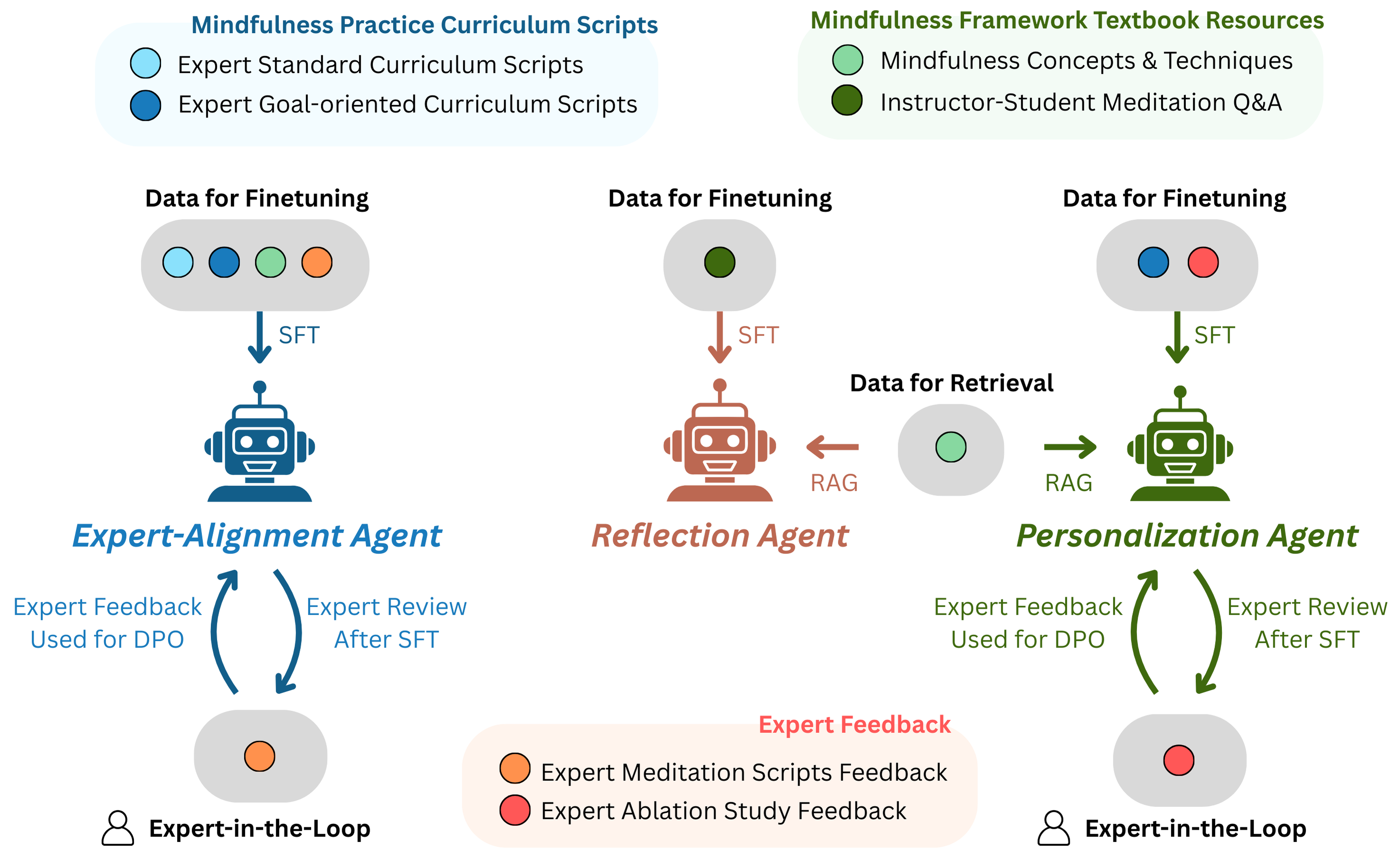}
    \caption{LLM Training and Retrieval Pipelines for Each Agent.
    \textbf{Left)} For the Expert-Alignment Agent, mindfulness practice curriculum scripts and UM concepts are used for supervised finetuning (SFT). Experts' edits and feedback on the intermediate generated scripts are included as additional data to further tune the model via both SFT and direct preference optimization (DPO);
    \textbf{Middle)} For the Reflection Agent, meditation Q\&A dialogues between instructors and students are used for SFT to emulate reflective conversations. And UM concepts are used for retrieval-augmented generation (RAG);
    \textbf{Right)} For the Personalization Agent, the SFT and DPO pipeline is similar to that of the Expert-Alignment Agent, except that standard curriculum scripts and mindfulness concepts are no longer included to better focus on goal-oriented scripts during the finetuning. Meanwhile, UM concepts are used for RAG to ensure content reliability.
    }
    \Description{Figure 2 illustrates the LLM training and retrieval pipelines for three agents in the MindfulAgents system. On the left, the Expert-Alignment Agent is trained with mindfulness practice curriculum scripts using supervised finetuning (SFT), with iterative expert review and feedback loops applied through direct preference optimization (DPO). In the center, the Reflection Agent is finetuned on Q\&A dialogue data from meditation instructors and students, using retrieval-augmented generation (RAG) for reflective conversations, and further guided by expert feedback. On the right, the Personalization Agent follows a similar pipeline to the Expert-Alignment Agent, but excludes standard curriculum scripts during finetuning, instead leveraging goal-oriented scripts and UM concepts via RAG. Across all agents, expert-in-the-loop processes ensure alignment, reliability, and personalization of outputs.}
    \label{fig:multi-agent-data}
\end{figure*}

\subsubsection{Phase 1: Supervised fine-tuning (SFT)}
We leverage the materials from UM concept introduction and meditation scripts (see details in \textbf{DG1} in Sec.~\ref{sub:co-design:co-design}) to finetune a GPT-4.1 model. The materials include both single-turn (concept introduction) and multi-turn (UM meditation scripts typically have the user select from 2-3 options, and their choice determines which predefined branch of the script they receive) training samples.
Finetuned on these resources, the Expert-Alignment Agent generated an initial batch of meditation guidance scripts varying along two key dimensions: \textit{duration} and \textit{goal}. Following the experts' suggestion, we standardized the duration at three levels (short 5-min, medium 10-min, and long 15-min) to fit into users' daily schedules. 
To preserve the style of UM user interactions during the guidance, we incorporated one interaction point into medium and long scripts, while short scripts did not contain interaction.
Moreover, in the generation pipeline, we further added LLM-based auto-checkers and correctors if errors in format, ending, and definition are detected in the script.

\subsubsection{Phase 2: Incorporation of Experts' Feedback}
To further refine and improve the model, experts reviewed a subset of the preliminary scripts and provided detailed edits. This human-in-the-loop process served multiple purposes: ensuring adherence to the UM framework, correcting subtle inaccuracies or phrasing that could lead to confusion, and refining tone to better match the supportive style of human instructors. 
We incorporated this feedback as training samples of both SFT and direct preference optimization (DPO)~\cite{rafailovDirectPreferenceOptimization2024} to further refine the model.
Specifically, expert-edited scripts were used directly as high-quality SFT data, and also served as the \textit{preferred} samples in DPO training. In contrast, the original LLM-generated drafts before expert edits were designated as the \textit{unpreferred} samples, creating structured comparison pairs for preference learning. We further augmented the DPO unpreferred pool by deliberately injecting malformed or improperly formatted variants of the generated scripts. This could reinforce the model's ability to distinguish well-formed from poor outputs \cite{katz-samuels_evolutionary_2024, paceWestofNSyntheticPreferences2024}.

\subsubsection{Phase 3: Content Validation by Experts}
After the two phases of finetuning, experts conducted a lightweight approval process to validate the completed meditation scripts. This step acted as a final safeguard, ensuring that only expert-aligned and error-free content entered the production corpus. All approved templates were then stored in the vector database, enabling efficient semantic retrieval at runtime.
To handle cases where a user's request does not closely align with any existing template, we also included a \emph{general template} designed to provide safe, flexible guidance that can accommodate a broad range of scenarios.

Overall, this three-phase design not only guarantees safety and reliability of the guidance generation, but also establishes an extensible pipeline: as new user needs or mindfulness scenarios evolve, additional expert-approved scripts can be seamlessly integrated into the database without retraining the model. In this way, the system balances stability with adaptability, maintaining expert oversight while enabling continual expansion.

\subsection{Reflection Agent: Reflection Prior to Meditation}
\label{sub:system:reflective_agent}

To incorporate reflection into the pipeline (\textbf{DG2}), we developed a chatbot that
encourages users to reflect on their current state, review insights from past meditation sessions, and reinforce their knowledge of meditation terminology. We selected GPT-4.1 nano as the chatbot backbone model to optimize the real-time, fast interaction for this lightweight activity. 
The model was fine-tuned on over 30 hours of meditation lesson dialogues using multi-turn SFT to emulate authentic interactions between meditation instructors and students (\textbf{DG1}), as shown in the middle of Fig.~\ref{fig:multi-agent-data}.
Building on this finetuned model, the agent provides three key functions for user reflection, each enriched with targeted contextual inputs.

\subsubsection{Reflecting on present}
This function leverages session-specific inputs (mood, goal) to help users articulate their current contextual details.
The agent initiates a brief, conversational check-in, prompting the user with open-ended questions about their current state (\eg a potential prompt when users select negative mood, ``I'm really sorry to hear you're not feeling well. If you'd like, I'm here to listen—would you like to share a bit more about what’s been going on?'').
This process encourages users to ground themselves in the present moment before formal practice begins and provides rich, real-time context that the Personalization Agent (Sec.~\ref{sub:system:personalization_agent}) uses to tailor the upcoming session.

\subsubsection{Reflecting on past sessions}
This function draws on historical meditation sessions, prompting users to revisit prior experiences and recognize progress over time.
Specifically, we used a vector database to store and search for the most relevant historical sessions. For example, the agent might ask, ``Last time, you were working with feelings of restlessness. Is that still present for you today?'' or ``What was one insight you took from your previous practice?''
This reflective loop helps reinforce learning, fosters a sense of a continuous personal journey, and allows the system to build upon previously covered ground.

\subsubsection{Reviewing meditation terms}
This function integrates a retrieval-augmented generation (RAG) pipeline over a vector database of UM concepts (see Fig.~\ref{fig:multi-agent-data} middle), enabling more accurate, context-sensitive explanations of key terminology, allowing users to ask for clarification on concepts like ``Equanimity'' or ``See, Hear, Feel''.

Overall, these functions work together to provide an opportunity to foster reflection before meditation to deepen engagement and strengthen continuity in meditation practice.%

\begin{table*}[b!]
\centering
\renewcommand{\arraystretch}{1.5}
\caption{Ablation Studies Comparing Input Structure of the Personalization Agent.
Average expert ratings across alignment, personalization, and overall quality (mean $\pm$ standard deviation across simulated persona). Scores of the same condition in two studies may vary as the content was re-generated in each ablation comparison.
Wilcoxon signed-rank tests were conducted to compare the scores. $r$ = rank-biserial correlation effect size. Statistical significance: $p < 0.05^{*}$, $p < 0.01^{**}$, $p < 0.001^{***}$.
}
\label{tab:ablation}
\resizebox{\textwidth}{!}{%
\begin{tabular}{l p{10.8cm} c p{3cm}}
\toprule
\textbf{\makecell{Ablation \\ Study}} & \textbf{Model \& Prompt Setups} & \textbf{\makecell{Expert Ratings \\ (1--10)}} & \textbf{\makecell{Comparison Stats}} \\
\midrule
\multirow{2}{*}{Study 1} & (A) \textcolor{black}{Base GPT model with Role Prompt} & $5.72 \pm 1.21$ & \multirow{2}{*}{\makecell{W=6.00, $p < 0.001^{***}$ \\ effect size $r=0.71$}} \\
& (B) \textcolor{black}{Finetuned Agent} & $\mathbf{7.76 \pm 0.88}$ & \\
\midrule
\multirow{2}{*}{Study 2} & (B) \textcolor{black}{Finetuned Agent} & $6.73 \pm 1.07$ & \multirow{2}{*}{\makecell{W=15.00, $p < 0.001^{***}$ \\ effect size $r=0.80$}} \\
& (C) \textcolor{black}{Finetuned Agent} \textcolor{black}{+ Technique Refresher} & $\mathbf{7.91 \pm 1.07}$ & \\
\midrule
\multirow{2}{*}{Study 3} & (C) \textcolor{black}{Finetuned Agent} \textcolor{black}{+ Technique Refresher} & $7.89 \pm 0.94$ & \multirow{2}{*}{\makecell{W=17.00, $p < 0.05^{*}$ \\ effect size $r=0.38$}} \\
& (D) \textcolor{black}{Finetuned Agent} \textcolor{black}{+ Technique Refresher + Reflection Content} & $\mathbf{8.51 \pm 0.72}$ & \\
\midrule
\multirow{2}{*}{Study 4} & (D) \textcolor{black}{Finetuned Agent} \textcolor{black}{+ Technique Refresher + Reflection Content} & $8.46 \pm 0.50$ & \multirow{2}{*}{\makecell{W=89.50, $p = 0.562$ \\ effect size $r=0.10$}} \\
& (E)\xspace\ \textcolor{black}{\xspace Finetuned Agent} \textcolor{black}{+ Technique Refresher + Reflection Content + Recent Summary} & $8.54 \pm 0.64$ & \\
\bottomrule
\end{tabular}
}
\end{table*}

\subsection{Personalization Agent: Personalized Meditation Guidance}
\label{sub:system:personalization_agent}

The key purpose of \projectname is to create reliable and personalized meditation guidance (\textbf{DG3}). Building on top of the Expert-Alignment Agent and Reflection Agent, we are ready to create the central personalization piece of the multi-agent system.

We carefully design two important aspects of the Personalization Agent: backbone model finetuning and the structure of the model inputs.

\subsubsection{Model Finetuning}
\label{subsub:system:personalization_agent:finetune}
We used GPT-4.1 mini to balance the generation capability and response time.
Although the Expert-Alignment Agent already provides safe meditation templates, the Personalization Agent should still be aware of mindfulness concepts and avoid hallucination. Therefore, we finetuned the GPT-4.1 mini with similar phases of the Expert-Alignment Agent (see Sec.~\ref{sub:system:expert_aligned_agent}). The main distinction was that only goal-oriented scripts were used among the curriculum scripts (\textbf{DG1}) during the fine-tuning, as these were most directly aligned with the final outputs of the personalization pipeline.
We also incorporated rule-based auto-checkers and correctors to catch format and ending errors in the script.

\subsubsection{Agent Input Prompt Design via An Ablation Study}

In addition to the model finetuning, an appropriate agent input structure that contains user inputs (mood, goal, duration, and technique, see details in Sec.~\ref{sub:system:interface}) is also important for personalization.
Given the different combinations of these components,
we conducted a series of ablation studies to identify the optimal prompt structure for the Personalization Agent.
Each ablation study was a pairwise comparison between two input structures, where we added one additional component at each round (see Table~\ref{tab:ablation}). For each pair, we generated matched sets of meditation scripts using identical simulated 12 user personas and the same session inputs (see Appendix Tab.~\ref{tab:user_persona} for the list of persona and their profile information).
To avoid cross-study leakage, scripts were re-generated for every ablation.
Two human experts, who were blinded to condition, rated each script from three dimensions on a 10-point Likert scale: \emph{Alignment} (adherence to the UM framework, including concepts, techniques, and terminology), \emph{Personalization} (the extent of personalization and relevance to the user’s persona and needs), and \emph{Overall Quality} (holistic meditation script quality).

\begin{figure*}[hb!]
    \centering
    \includegraphics[width=1\linewidth]{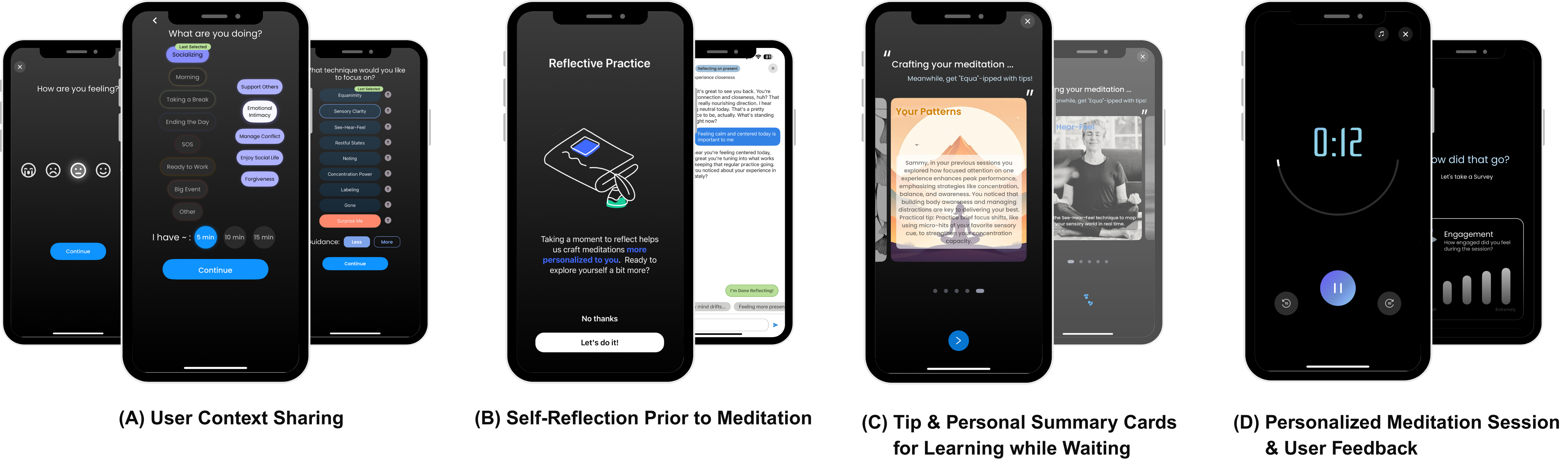}
    \caption{The User Flow across \projectname Interface.
    \textbf{(A)} The user starts by sharing their mood, goals, as well as preferred mindfulness technique, duration, and level of guidance.
    \textbf{(B)} The user then has the option to interact with the Reflection Agent to share their thoughts and experience.
    \textbf{(C)} The user could read the tip and personal summary cards while waiting for the personalized meditation content and audio being generated.
    \textbf{(D)} The user enters the meditation experience with an audio player and reports feedback at the end of the meditation session.
    }
    \Description{Figure 3 shows the user flow across the MindfulAgents mobile interface. From left to right: (A) Users begin by sharing context, including mood, goals, meditation technique, duration, and guidance preference. (B) Before meditation, they can engage in self-reflection through a chatbot interface. (C) While waiting, users view personalized tips and summary cards for learning and reflection. (D) Finally, they participate in a personalized meditation session via an audio player and provide feedback afterward.}
    \label{fig:personalized_interface}
\end{figure*}

\begin{enumerate}
     
\item \textit{Study 1—Model.}
We first isolated the effect of model finetuning by comparing a base GPT model against our finetuned one (Sec.~\ref{subsub:system:personalization_agent:finetune}).
The first row in Table~\ref{tab:ablation} showed that our finetuning process for the Personalization Agent significantly improved the content quality (Wilcoxon signed-rank test $p<0.001$, rank-biserial correlation effect size $r=0.71$, the same below).
\item \textit{Study 2—Technique Refresher.}
We next enhanced the prompt with a brief ``technique refresher'' for the user-selected mindfulness technique, retrieved from the RAG pipeline. The refresher included the technique's definition and key steps, and was injected into the prompt whenever a technique was chosen. This addresses a common model failure where it either omits or hallucinates technique details, thereby stabilizing terminology and making the instructions more precise and actionable (the 2nd row in Table~\ref{tab:ablation}, $p<0.001, r=0.80$).
\item \textit{Study 3—Reflection Content.}
We further augmented the prompt with reflection content to increase contextual richness and personal relevance. Specifically, we included the chat history with the Reflection Agent and a summary of the three most related prior sessions given the current inputs. These inputs strengthen the continuity across sessions, highlight user progress, and support the discovery of meaningful trends in practice, leading to higher ratings (the 3rd row in Table~\ref{tab:ablation}, $p<0.05, r=0.38$)
\item \textit{Study 4—Recent Summary.}
Finally, we appended a compact summary of short-term trends (daily check-in ratings for sleep, mood, and focus) together with a summary of the three most recent sessions.
Due to significant conceptual overlap with the Reflection Content, we observed saturation of performance improvement (the last row in Table~\ref{tab:ablation}, $p=0.56, r=0.10$).
\end{enumerate}

Balancing benefit and brevity, we adopt the configuration comprising the user profile, the \emph{Technique Refresher} and \emph{Reflection Content}. This configuration yielded consistent, practically meaningful gains while keeping prompts succinct.

\subsubsection{Finalized Personalization Agent Input Structure}

{Our final Personalization Agent operates on top of expert-aligned scripts and conditions its adaptations on all six dimensions identified in DG3: (1) Mood, (2) User Goals, (3) Technique, (4) Duration, (5) Level of Guidance, (6) Practice History. The Expert-Alignment Agent first produces a bank of expert-validated script variants (from Sec.~\ref{sub:system:expert_aligned_agent}) indexed by (2) \textit{user goal}, (4) \textit{desired session duration}, and (5) \textit{level of guidance}. The Personalization Agent then selects an appropriate variants and further personalizes it in real time using (1) the user’s current \textit{mood}, (3) preferred \textit{technique}, and (6) \textit{practice history}, informed by structured inputs and ongoing conversations with the Reflection Agent (from Sec.~\ref{sub:system:reflective_agent}). (1) \textit{Mood} is derived from explicit user input and “reflecting on present” exchanges, shaping the tone and normalization language of the script and weaving in concrete situations the user has shared. (3) \textit{Technique} is based on user choice and a brief \textit{technique refresher} retrieved from a curated library of Unified Mindfulness techniques and prior “reviewing meditation terms” interactions, determining which exercise is introduced and how it is framed given the user’s questions and points of confusion. (6) \textit{Practice history} is represented via \textit{reflection content} that combines selected chat history and summaries from “reflecting on past sessions,” allowing the system to reference previous challenges and progress and scaffold continuity across sessions. Together, this multi-agent pipeline couples expert-aligned content with coach-like pre-practice reflection and fine-grained conditioning on all six DG3 dimensions. To our knowledge, this moves beyond typical meditation apps and generic LLM-based coaches, which usually focus on routing users to pre-authored meditations rather than real-time script-level personalization.}

\subsection{\projectname Interface Design}
\label{sub:system:interface}

We developed a mobile application to instantiate \projectname with an interface.
As illustrated in Fig.~\ref{fig:personalized_interface}(A), the user journey begins with a series of simple personalization screens designed for quick input. Users first select their current mood. Then, they choose a meditation goal and a preferred session duration (5, 10, or 15 minutes). They also select a mindfulness technique and set their desired level of guidance (``More/Less'' toggle). To support learning, each technique is accompanied by an information icon that can provide a brief definition when clicked. To promote continuity in future sessions, the interface dynamically adapts the order of goals and techniques based on the user's practice history.

Following these initial selections, users are given the option to engage with the Reflection Agent, as shown in  Fig.~\ref{fig:personalized_interface}(B).
If they choose to proceed, the conversation starts by default with reflecting on their present state. They can switch to discussing past sessions or reviewing UM terms.
To ensure user agency, they can end the conversation at any point or skip the reflection stage entirely and proceed directly to the meditation session.

Once the inputs are complete, the Personalization Agent takes the input from the Expert-Aligment Agent and the Reflection Agent and begins generating the guided meditation content and audio.
While the content is being prepared, a Sliding Cards interface appears (Fig.~\ref{fig:personalized_interface}(C)), transforming the brief waiting period into a learning opportunity. These cards present short, digestible mindfulness tips as well as personalized summaries and reflections from past sessions. Finally, the user enters the core meditation experience, which features a clean audio player interface, as shown in Fig.~\ref{fig:personalized_interface}(D).
Upon completion, the app prompts them to provide feedback on the session, which helps refine future personalization and contributes to their practice history.

\section{Formative Lab Study}

\label{sec:us1}
{In this formative lab study, we explored how users experienced different components of \projectname---expert alignment, detailed personalization, and reflective interaction---in a controlled setting prior to a longer-term field deployment. Our goal was to characterize momentary user experience across system variants and to collect early feedback to inform the design of the subsequent field deployment study.}

\begin{figure*}[t]
    \centering
    \includegraphics[width=0.85\linewidth, height=0.30\textheight, keepaspectratio]{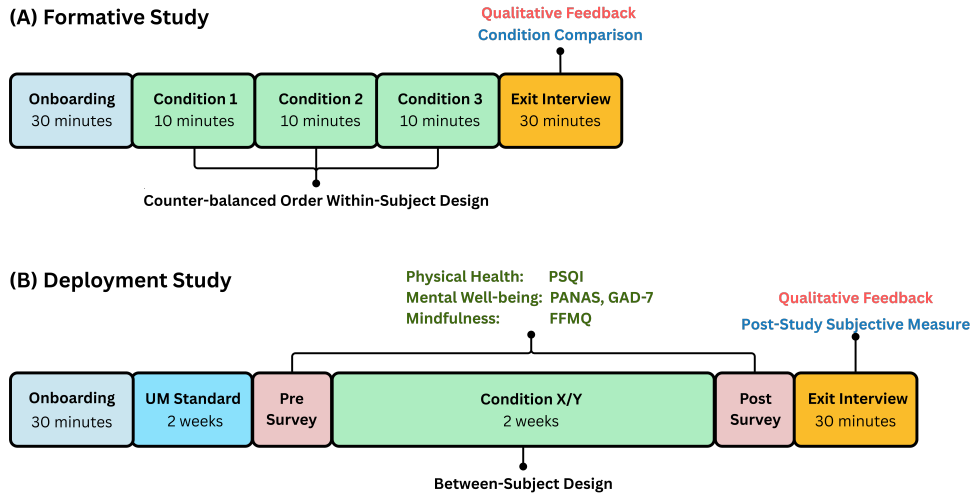}
    \caption{Overview of the {Formative Lab Study and Field Deployment Study}.
    \textbf{(A)} Formative Study used a within-subject design (N=13) to compare user experience across three ablation conditions: StaticAgent (\projectname without personalization or reflection), PersonalAgents (\projectname without reflection), and the full \projectname system.
    \textbf{(B)} Deployment Study used a between-subjects design (N=62) to compare StaticAgent and \projectname over a four-week in-the-wild deployment, focusing on long-term engagement and changes in mindfulness and broader well-being.}
    \Description{Figure 4 shows study designs in flowchart format for two user studies of MindfulAgents. Shown at the top, (A) Formative Study: within-subject design with onboarding (30 min), three 10-minute conditions in counterbalanced order, and a 30-minute exit interview for qualitative feedback and condition comparison. Shown at the bottom, (B) Deployment Study: between-subject design with onboarding (30 min), a two-week UM Standard phase, pre-survey, two-week experimental condition (X or Y), post-survey, and a 30-minute exit interview. Measures include physical health (PSQI), mental well-being (PANAS, GAD-7), and mindfulness (FFMQ).}
    \label{fig:us_design}
\end{figure*}

\subsection{Baseline Conditions}
As \projectname is a multi-agent system, we compare it against two baseline agentic systems in an ablation manner. 

\subsubsection{Baseline 1: StaticAgent (\projectname w/o Personalization and Reflection)}
\label{baseline_condition} 

{In this condition, participants engaged with the same app but in a static, non-personalized content mode. StaticAgent includes only the Expert-Alignment Agent, which provides reliable meditation guidance scripts drawn from a pre-generated library. For this study, the system used the participant’s selected \textit{goal} to index into this library and select a pre-scripted meditation; other fields were held constant across sessions and did not change the script content. We intentionally limited indexing to a goal to approximate a minimal personalization baseline commonly found in commercial meditation apps, where users typically choose from scripts labeled by high-level goals (e.g., “sleep,” “stress relief”) rather than fully tailored guidance.}
To control for potential confounds of the interface, we retained the same emotion and goal selection interface. Moreover, to control for confounds of the content generation, we implemented a pseudo-generation delay, ensuring participants waited a similar amount of time as in the other conditions, even though no real-time AI generation occurred.

\subsubsection{Baseline 2: PersonalAgents (\projectname w/o Reflection)}
\label{base_personalization_condition}

{This condition builds on the first baseline and employs both the Expert-Alignment Agent and the Personalization Agent. The Expert-Alignment Agent uses the user’s selected \textit{goal}, \textit{duration}, and \textit{guidance preference} to retrieve and structure meditation scripts, while the Personalization Agent further tailors these scripts based on \textit{mood}, chosen \textit{technique}, and practice history. In this way, PersonalAgents leverages the same structured user inputs as \projectname to generate dynamic, contextually relevant, expert-aligned meditation content. However, PersonalAgents does not include the Reflection Agent, so personalization in this baseline does not include the fine-grained, free-text reflective input about the user’s current situation, confusions about concepts, or perceived progression in their meditation journey.}

\subsection{Study Design \& Procedure}

{We adopted a within-subject design for the formative study where participants interacted with all three systems (StaticAgent, PersonalAgents, and \projectname). This design increased sensitivity to differences in perceived user experience and allowed us to gather richer comparative feedback. The order of conditions was counterbalanced across participants to mitigate ordering effects.
}
This study has been approved by Columbia University Institutional Review Board (IRB; Protocol \#AAAV7750) and conducted in accordance with IRB regulations. Participants provided informed consent; data were stored and analyzed in de-identified form.

The session began with a 30-minute onboarding process where participants were briefed on the study and signed a consent form.
They also watched a short tutorial video to ensure basic understanding of UM concepts before installing the application on their devices.
They then engaged with each of the three intervention techniques for approximately 10 minutes, following the procedure illustrated in Fig. \ref{fig:us_design}(A).
After interacting with all conditions, participants completed a short questionnaire to provide in-session user experience ratings on five aspects on a 1-5 Likert scale, including in-session engagement, level of personalization, deepening self-awareness, stress reduction, and willingness for future use, {details provided in Table~\ref{tab:formative_measures} in the Appendix.}
In addition, they also ranked the three conditions according to overall preference, helpfulness, and potential for future use.
The session concluded with a 20 to 30-minute semi-structured interview to gather qualitative feedback on their experience with each technique. Each participant was compensated \$25 for their time.

\subsection{Participants}
We recruited participants by posting digital flyers on online communities interested in meditation (\eg Facebook groups, Reddit). We used a screening questionnaire to collect basic demographics and experience with meditation. 14 participants were initially recruited and one dropped out, resulting in a final sample of $N=13$.
The sample included 8 males and 5 females, with a mean age of 27.5 ($SD=3.7$). Regarding prior experience, 7 participants reported having tried meditation once or twice, 4 practice occasionally, and 2 had never practiced.

\subsection{Results}

\subsubsection{In-Session User Experience Rating}
We first examined in-session user experience ratings across all conditions on all metrics. As summarized in Fig.~\ref{fig:us1_results}(A), Friedman tests revealed significant differences for in-session engagement ($\chi^{2}(2) = 8.97$, $p = 0.011$), level of personalization ($\chi^{2}(2) = 7.74$, $p = 0.021$), deepening self-awareness ($\chi^{2}(2) = 8.58$, $p = 0.014$), and stress reduction ($\chi^{2}(2) = 7.85$, $p = 0.020$).
Post-hoc pairwise comparisons with Wilcoxon signed-rank tests with Holm–Bonferroni correction show that both PersonalAgents and \projectname achieved significantly higher in-session engagement than StaticAgent ($r=0.82, 0.82$, respectively), and that \projectname outperforms StaticAgent on deepening self-awareness ($r=0.91$) and stress reduction ($r=0.89$). Overall, although there is no pairwise significant difference observed between PersonalAgents and \projectname, the advantage of \projectname is reflected in greater improvement over the StaticAgent baseline.

\subsubsection{User Preference Rankings}
\begin{figure*}[t]
    \centering
    \includegraphics[
      width=\linewidth,
      keepaspectratio
    ]{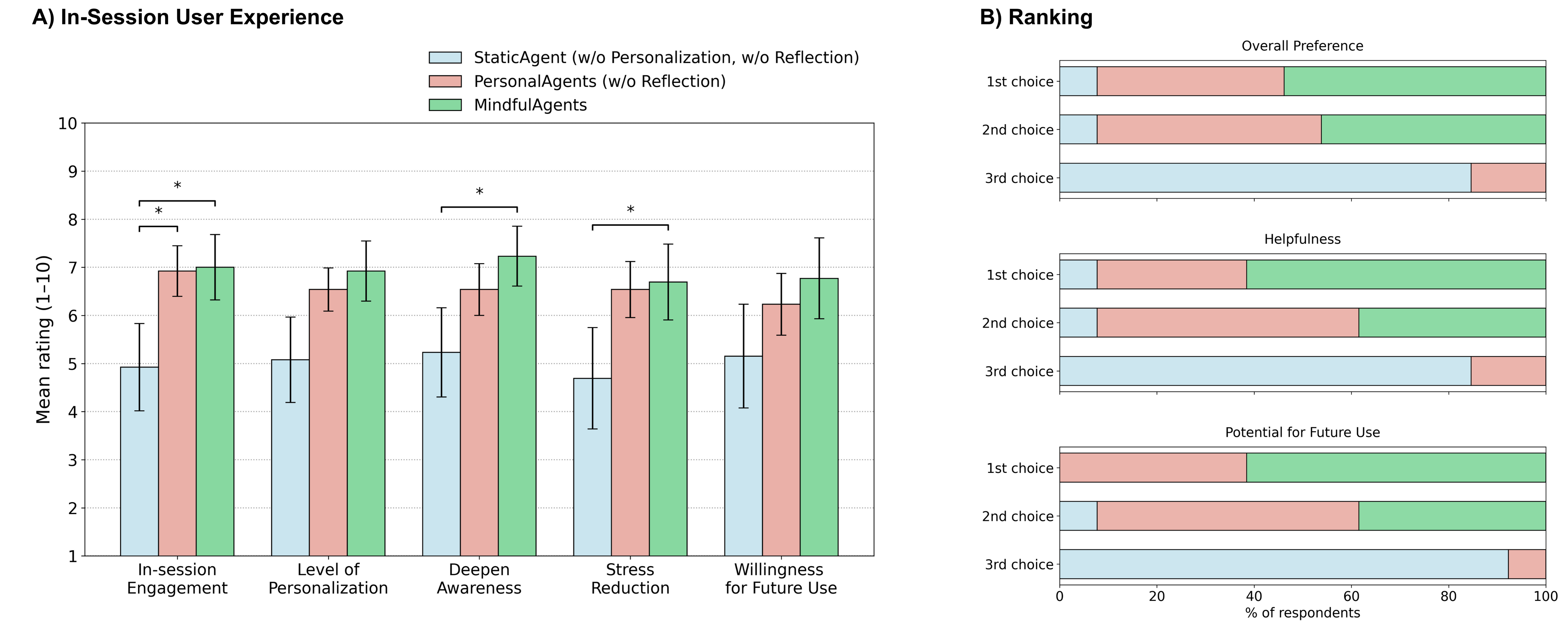}
    \caption{{Formative Study} Results.
    \textbf{(A)} Scores across five user experience metrics demonstrate that \projectname has the overall best performance. Error bars indicate standard error.
    \textbf{(B)} Preference rankings show consistent findings. \projectname received the highest ranking overall, followed by PersonalAgents, and then StaticAgent.
    }
    \Description{Figure 5 presents results from the Formative Lab Study. On the left, (A) Bar chart of in-session user experience ratings (1–10) across five metrics: in-session engagement, level of personalization, deepen awareness, stress reduction, and willingness for future use. MindfulAgents show consistently higher scores than PersonalAgents and StaticAgent, with significant improvements in engagement, awareness, and stress reduction (p < 0.05). Error bars show standard error. On the right, (B) Stacked bar charts of preference rankings across overall preference, helpfulness, and potential for future use. MindfulAgents is most frequently ranked first, followed by PersonalAgents, then StaticAgent.}
    \label{fig:us1_results}
\end{figure*}

Ranking results lead to a consistent conclusion among overall preference, helpfulness, and potential for future use.
Friedman tests revealed significant differences across all three criteria, as shown in Fig.~\ref{fig:us1_results}(B). For overall preference ($\chi^{2}(2)=12.15$, $p=0.002$), 84.6\% of participants ranking StaticAgent the last.
In contrast, 38.5\% of participants selected PersonalAgents and 53.8\% selected \projectname as their top choice, indicating a clear preference for our system. This trend was even more pronounced for the helpfulness and the willingness for future use ($\chi^{2}(2)=12.92,p=0.002$ and $\chi^{2}(2)=17.23,p<0.001$), where StaticAgent was almost universally ranked last (84.6/92.3\%), and the \projectname was the most appealing (both 61.5\%).
Overall, these results highlight a strong and consistent preference for \projectname over the static baseline, with PersonalAgents also showing promising results.
The clear advantage of personalization also suggests that, although both reflection and personalization play a critical role in improving the mindfulness experience, the importance of personalization may be higher than that of reflection.

\subsubsection{Qualitative Feedback}
We conducted a brief content analysis of the interview notes. Participants' comments provide additional insights. Participants often described the \projectname as emotionally resonant and relevant. For instance, P07 appreciated how narrating their personal experience beforehand \textit{``gave me some resting place so I can make this session more engaged and more focused''}. In contrast, StaticAgent was described as ``generic'' and ``distracting'', mainly due to the content being overly broad and irrelevant. This is supported by prior research showing the low engagement with meditation apps despite providing structured and reliable meditation content. Meanwhile, participants also provided nuanced feedback on the added value of the Reflection Agent. P12 preferred the \projectname condition more because \textit{``I could tell the session was personalized, um, based on my answers [from reflection].''} Similarly, P07 emphasized the Reflection Agent as a critical differentiator, noting, \textit{``In some sense, I’m talking to someone else… I can discuss my own scenario and situation to the chat window... I think that’s really one differentiator as well.''}. 
Although our post-hoc analysis showed only marginal benefits of \projectname over PersonalAgents, the qualitative insights suggest that reflection may offer a unique pathway to deeper engagement, informing the design of our long-term field study in Sec.~\ref{sec:us2}.

\section{In-the-Wild Deployment Study}
\label{sec:us2}

{Our formative lab study suggested that participants preferred the personalized conditions over the static baseline and that adding reflection atop structured personalization was a promising direction. For the deployment study, our focus therefore shifted from component-level ablation to evaluating the real-world impact of the system. We centered the study on the most practically relevant comparison: the StaticAgent baseline, which approximates most state-of-the-art LLM-based mindfulness solutions available on the market, versus the full \projectname system implementing the complete design. Building on these insights, we conducted an IRB-approved (Protocol \#AAAV7750) four-week between-subjects deployment study comparing StaticAgent and \projectname to assess longer-term engagement and self-reported outcomes.}

\subsection{Study Design \& Procedure}
\label{sec:us2:study_design}

We conducted a four-week {deployment} study to evaluate the long-term engagement, user experience, and changes in mindfulness and broader well-being after using our system. 
The four-week study began with a 30-minute onboarding session where participants were briefed on the study, signed consent forms, and installed the application. The study was then conducted in two distinct phases.

\textbf{Phase 1: Standardized Curriculum (Weeks 1-2).} For the first two weeks, all participants engaged with a standardized UM curriculum. To ensure a comparable baseline understanding of the techniques, participants were required to complete at least 11 of the 14 training units before advancing to the next phase.

\textbf{Phase 2: Free Practice (Weeks 3-4).} Following the training period, a brief follow-up session was held to introduce participants to the features of their assigned condition. For the following two weeks, participants entered a free-practice period and were randomly assigned to either of the two conditions: (1) StaticAgent Condition, (2) \projectname Condition. They were informed that they could practice meditation with the app at their own discretion, and that their compensation was not linked to the number of meditation sessions. Participants in both conditions would receive automatic daily reminder notifications from the app. Throughout this phase, we logged user behaviors with the application.

We collected four types of data across the four-week study, including:
\begin{itemize}
    \item \textbf{Interaction Logs}. During the two-week free-practice phase, we automatically logged participants' interaction behaviors with the application.

    \item \textbf{Behavioral Surveys}. At the beginning (pre-) and the end (post-) of the Free Practice period, participants completed surveys assessing physical health (Pittsburgh Sleep Quality Index, PSQI~\cite{buyssePittsburghSleepQuality1989}), mental well-being (Positive and Negative Affect Schedule - Short Form~\cite{mackinnonShortFormPositive1999}, PANAS-SF; Generalized Anxiety Disorder~\cite{spitzerBriefMeasureAssessing2006}, GAD-7), and mindfulness (Five Facet Mindfulness Questionnaire–Short Form~\cite{bohlmeijerFiveFacetMindfulness2011}, FFMQ-SF).
    
    \item \textbf{Final User Satisfaction Survey}. At the conclusion of the study, a final questionnaire was administered to gather quantitative ratings on several personalization aspects:  user experience (goal alignment and inclusiveness), meditation content quality (guidance \& pace and concept clarity), and System Usability Scale (SUS)~\cite{brook_sus_1995}. 
    
    \item \textbf{Exit Interview}. Semi-structured interviews were conducted at the end of the study to collect in-depth qualitative feedback on participants' experiences.
\end{itemize}

\subsection{Participants}
Similar to our Formative Study, we recruited participants by sending a digital poster on online communities and used a screening questionnaire to filter out ineligible users.
91 users were initially onboarded and randomly assigned to one of the agent conditions. 24 of them completed less than 11 lessons in Phase 1, and 3 had data connection or access issues. 
64 participants entered Phase 1 and 2 of them dropped out voluntarily, yielding a final sample of 62 with 29 participants in \projectname condition and 33 participants in the StaticAgent condition.
The final sample included 33 females, 27 males, and 2 non-binary individuals, with a mean age of 44.0 ($SD=14.0$).
Most of the participants had prior experience with meditation (83.9\%) and AI tools (79.1\%). Participants are compensated up to \$60, depending on their level of involvement in the study phases.

\subsection{Results}
Across the 4-week study, participants completed a total of 1,766 structured curriculum sessions in Phase 1 and 704 free meditation sessions in Phase 2 (279 for StaticAgent and 425 for \projectname). 
Participants actively integrated meditations into their daily routines, most often pursuing broader goals such as stress relief (29.8\%), winding down at the end of the day (28.2\%), and cultivating a sense of fulfillment (19.7\%).
We present both quantitative analyses of usage patterns and questionnaires and qualitative insights from interviews.

\begin{figure}[ht!]
    \centering
    \includegraphics[width=1\linewidth]{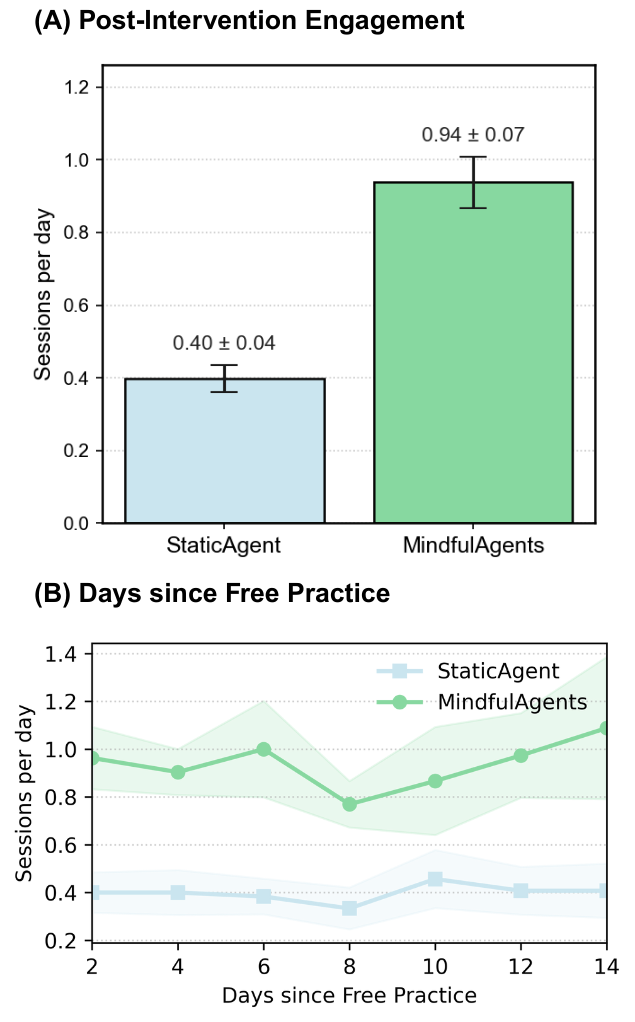}
    \caption{{Deployment Study} Results on Long-Term Engagement during the Free Practice Phase. \textbf{(A)} \projectname achieved a higher post-intervention engagement than that of StaticAgent, and \textbf{(B) }this advantage sustains over time throughout the free practice period.}
    \Description{Figure 6 compares post-intervention engagement between StaticAgent and MindfulAgents. On the left, (A) Bar chart shows mean daily sessions during the free meditation stage: MindfulAgents users practiced significantly more (0.94 ± 0.07) than StaticAgent users (0.40 ± 0.04). On the right, (B) Line graph shows sessions per day across 14 days: MindfulAgents maintained consistently higher engagement, with rates increasing over time, while StaticAgent remained low and flat.}
    \label{fig:us2_key_results}
\end{figure}

\subsubsection{{Post-Intervention Engagement}}
One of the direct indicators of the long-term engagement of \projectname is how many free meditation sessions the users continue to engage in after the standard meditation curriculum. {We defined \textit{post-intervention engagement} as the average number of sessions per day during the two-week free-practice phase, capturing the intensity of practice rather than a binary dropout status.
For each participant $i$, we computed post-intervention engagement as the average number of meditation sessions completed per day during the two-week free-practice period:
\[
E_i = \frac{\text{\# sessions completed by participant } i}{\text{\# days in free-practice phase}}.
\]
Given the limited duration of the free-practice phase and the study's low attrition, we prioritized a frequency-based metric to capture the depth of engagement rather than simply tracking whether participants remained enrolled. We report the mean and standard deviation of $E_i$ within each condition. } As shown in Fig.~\ref{fig:us2_key_results}, the \projectname condition exhibited a significantly higher post-intervention engagement (Mean $= 0.94$, $SD=0.07$) than the StaticAgent baseline (Mean $= 0.40$, $SD=0.04$, two-sided Mann–Whitney U test $U=699, p=0.002$, $r =0.46$).

To further understand how engagement evolved over the two-week period, we fit a Generalized Linear Mixed Model (GLMM) to examine the effects of the condition (\textit{condition}), the passage of time (\textit{time}), and their interaction on the number of sessions users engaged in per day. Participant ID was included as a random intercept to account for individual differences. The results revealed a significant main effect of \textit{condition} ($b = 0.384$, $p = 0.015$), reinforcing that participants in the \projectname condition completed significantly more sessions per day. For the StaticAgent group, engagement levels remained stable, with no significant main effect of \textit{time} ($b = -0.009$, $p = 0.355$). Furthermore, we observed no significant \textit{condition} $\times$ \textit{time} interaction ($b = -0.005$, $p = 0.709$). %
Overall, these results suggest that personalized guidance in \projectname led to meaningfully higher daily meditation engagement, which remained throughout the entire period compared to the static baseline.

\subsubsection{Behavioral Survey}

\begin{figure*}[b]
    \centering
    \includegraphics[width=0.8\linewidth]{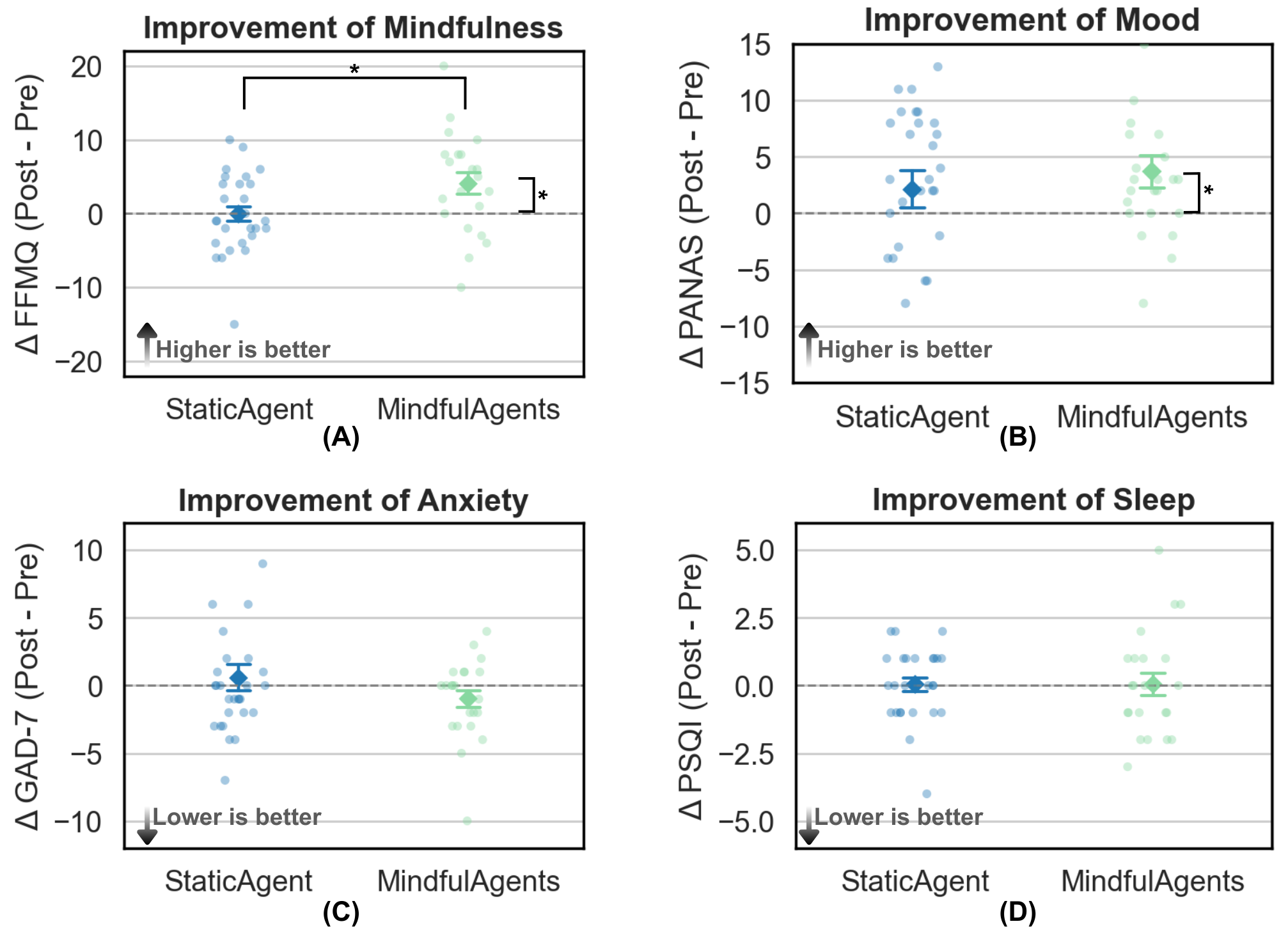}
    \caption{{Deployment Study} Results on the Changes in Behavioral Outcomes. Scores were computed as the delta between Post- and Pre-surveys. \textbf{(A)} mindfulness (FFMQ-SF), \textbf{(B)} mood (PANAS-SF), \textbf{(C)} anxiety (GAD-7), and \textbf{(D)} sleep quality (PSQI). \projectname shows better results in promoting mindfulness and mood level compared to the StaticAgent.}
    \Description{Figure 7 shows changes in behavioral outcomes, computed as post–pre survey deltas, comparing StaticAgent and MindfulAgents. (A) Mindfulness (FFMQ-SF): MindfulAgents produced significantly greater improvement than StaticAgent, with mean gains around +4. (B) Mood (PANAS-SF): MindfulAgents users showed significant within-condition improvement, though not significantly different from StaticAgent. (C) Anxiety (GAD-7): Both groups showed small, nonsignificant changes. (D) Sleep quality (PSQI): Minimal changes in both conditions. Significance markers indicate statistical tests, with only mindfulness showing between-condition differences favoring MindfulAgents.}
    \label{fig:us2_behavioral_results}
\end{figure*}

To complement the behavioral log data and measure the behavior change after Phase 2, for each participant, we calculate their delta score between the post- and pre-surveys. The results are summarized in Fig.~\ref{fig:us2_behavioral_results}. 

Wilcoxon Signed-Rank tests on each condition (compared to change$=0$) showed that mindfulness (FFMQ-SF, Mean$=4.09, SD=6.78, W=187.5,p = 0.006$) and mood (PANAS-SF, Mean$=3.70, SD=6.83, W=174.0,p = 0.005$) were improved after using \projectname, with statistical significance.
When comparing the two conditions,  a Mann–Whitney U test revealed a statistically significant difference  ($U = 424.5, p = 0.023, r = 0.322$) on mindfulness, with \projectname producing greater improvements in mindfulness.
Other comparisons did not show a significant difference on mood (PANAS-SF, $p = 0.917$), anxiety (GAD-7, $p = 0.587$), or sleep quality (PSQI, $p = 0.594$).

On the one hand, these results show that the higher long-term engagement with \projectname leads to promising improvements on mindfulness and mood level, outperforming the static baseline.
On the other hand, the modest changes on anxiety level and sleep behavior align with prior literature (see Sec.~\ref{sec:related_work:engagement}), which suggests that mindfulness interventions primarily enhance mindfulness and mood in the short term, whereas clinically meaningful improvements in anxiety or sleep typically require longer interventions over months or even years.

\subsubsection{User Experience Measure}

\begin{figure}
    \centering
    \includegraphics[width=0.99\linewidth]{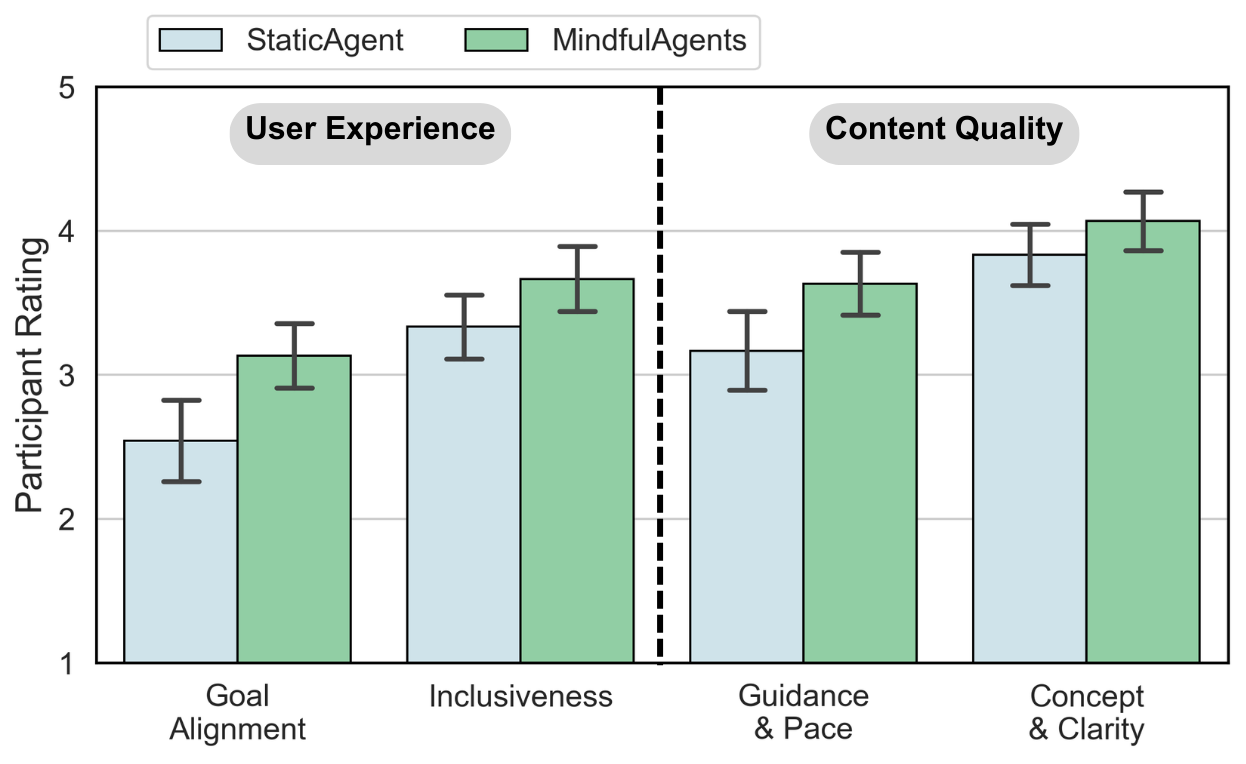}
    \caption{{Deployment Study} Results on Final User Satisfaction Scores. \projectname shows overall higher scores across all dimensions on user experience and content quality, although no statistical significance was observed.}
    \label{fig:us2_feedback}
    \Description{Figure 8 shows post-study participant satisfaction ratings of StaticAgent and MindfulAgents across two dimensions: User Experience (goal alignment, inclusiveness) and Content Quality (guidance and pace, concept and clarity). In all four categories, MindfulAgents received higher mean ratings than StaticAgent, with differences most pronounced for goal alignment and guidance and pace, although Mann-Whitney U tests did not show significance. Error bars indicate standard error.}
\end{figure}

Moreover, we collected participants' subjective evaluations of their satisfaction at the conclusion of the study.
Overall, participants rated \projectname more favorably than the \textit{StaticAgent} in all dimensions across goal alignment, inclusiveness, guidance \& pace, and concept clarity, as shown Fig.~\ref{fig:us2_feedback}, although Mann–Whitney U tests did not show significance.
In addition, participants rated ``good'' SUS scores on both  \projectname (Mean$=75.46, SD=19.07$) and StaticAgent (Mean$=75.45, SD=17.50$).

\subsubsection{Qualitative Feedback}
\label{results:qualitative_feedback}
We transcribed the interviews and conducted thematic analysis using open coding to find themes \cite{miles2014Quali} related to participants' perceptions towards \projectname and \textit{StaticAgent}. The first author led the analysis, creating codes and then broader themes via discussion with the other authors. Finally, the authors collaboratively reorganized the resulting themes. Participant ID was randomly assigned across 1-200.
In summary, participants highlighted three key advantages of \projectname.

\textbf{Personalizing sessions based on users' current and historical context made the guidance more relevant and applicable.}
Participants consistently reported that when the system acknowledged their specific context, \eg addressing their name, referencing past sessions, or tailoring content to their emotions, it fostered a stronger sense of being understood and deepened their engagement. 
For instance, P122 valued being able to \textit{``start the session by describing what's going on with me,''} finding that the system \textit{``wove that into the meditation''} in a way that felt new and engaging. 
P123 appreciated that the system offered a recap at the end of the session that remembered what they had worked on before. 
These comments underscore that when the system demonstrates contextual awareness,
it makes the experience more personally meaningful and deepens user engagement.

\textbf{Facilitating participants' reflection fostered a feeling of being understood and created a sense of companionship over time.}
Many users found that interacting with the reflection chatbot felt akin to journaling, helping them articulate emotions and reframe struggles in supportive ways (\eg P123, P156). P148 further emphasized that the AI \textit{``helped [me] do a better job of self-reflection, which is a great first step''}.
Similarly, P170 noted that discussing daily experiences with the chatbot and seeing them woven into the session created a feeling of being \textit{``supported when personalization worked correctly.''} 
Participants liked the Reflection Agent as it was like \textit{``a companion''} (P019). 
Together, these results suggest that the reflection has the potential to transform \projectname from a passive tool into an empathetic companion and make the subsequent meditation session more integrated and personally relevant. {Moreover, participants saw value in extending this interaction beyond the start of the session, suggesting that dynamic reflection placed in the middle or at the end of a module could help check comprehension and consolidate insights  (P019, P123).}

\textbf{Diverse meditation content drove sustained engagement.}
As the meditation content within \projectname was personalized to users' current and past contexts, it was inherently diverse and non-repetitive. This variety was a key driver of motivation. 
P007 explained that they felt excited to see \textit{``what's different''} each time they opened the app, framing novelty itself as a powerful incentive. 
Similarly, P110 appreciated the diversity of the generated content, observing that even small variations made the experience more enjoyable. 
Conversely, the lack of variety in StaticAgent diminished focus and interest. 
P059 called the control version \textit{``boring and annoying''} because selecting the same focus produced \textit{``the exact same text... over and over again.''} Likewise, P060 reflected that the static pipeline felt \textit{``very static''} and \textit{``template-like,''} offering little variability. Collectively, these comments suggest that novelty and subtle variation motivated sustained engagement, while the predictable repetition in the static condition risked causing disengagement.

\subsubsection{Suggestions for Improvement}
\label{subsub:us2:results:improvement}
Participants also provided some suggestions for improving our system.

\textbf{More Immersive and Smooth Experience.} Participants noted that a seamless flow was critical for an immersive meditation experience. Technical delays, however, often broke this immersion by creating frustrating uncertainty (P011) or preventing the session from starting at all (P036). Technical latency pulls users out of the meditative state by drawing attention to the system's mechanics (P058). To provide a more immersive and smooth meditation experience, the system must further reduce these delays and provide a more fluid experience.

\textbf{Enhance the human-likeness of voice to create a more natural and engaging user experience.} Participants suggested that a more natural voice for the AI meditation guidance could foster a stronger connection. While the AI voice was clear, its robotic tone sometimes undermined the experience, especially when personalizing content. One user noted that an unnatural pause when saying their name was a clear reminder \textit{``that it is AI''} (P094), breaking the sense of immersion. Although this is not the focus of our paper, improving prosody and reducing robotic artifacts could further increase the feeling of warmth and companionship.

\textbf{Balancing user efforts with personalization benefits.}
Feedback highlighted the need to reduce the manual effort required to receive personalized content. Some participants found the onboarding process \textit{``too long''} (P013), expressing an expectation that the AI should already \textit{"know what [they] want without so much explanation."}
When the setup feels burdensome (P011), it increases the perceived effort needed to get value from the system. This reveals a key design tension: the system must justify the user effort for data input by delivering a clearly superior and highly relevant personalized output.

{\textbf{Need for Social Calibration and Long-term Evolution.} Finally, participants highlighted that a companion must adapt not just to meditation preference, but to their cultural background and evolving expertise. Bilingual speakers (P025, P154), noted that \textit{''some standard phrasing felt alien and not common talk''}, suggesting a need for cultural linguistic sensitivity and potential enhancement with bilingual features. Others noted that the system sometimes lacked social calibration, for instance, the AI may repeatedly mention a specific shared event that seems forced and repetitive (P122, P138, P147). This suggests that effective personalization requires more than simple data retrieval; it demands sophisticated longitudinal memory and social calibration to ensure that the system understands not just \textit{what} to remember, but \textit{when} it is socially appropriate to reference it. Furthermore, users expressed that their needs changed as they gained experience. P093 explicitly described three stages of usage: \textit{''beginners need a teacher with structure, but experts need the app to "get out of their way"''}, eventually preferring self-mastery over constant instruction (P151). This suggests that the system's persona should not remain static, but evolve from a directive instructor to a supportive, low-friction tool.}

\textbf{Narratives of Becoming: Making Progress Tangible and Visible.}  
A repeated theme across interviews was the central role of progress tracking in sustaining motivation. Our system helped reflect progress by generating summaries of prior sessions to contextualize ongoing practice. For several, seeing continuity across sessions was already motivating: \textit{``I liked that it remembered what I worked on before—it felt much more relevant and added to my motivation to continue''} (P123). Beyond the current system, participants expressed a strong preference for more explicit forms of progress feedback so the system could be more integrated into a larger life narrative. As one participant put it, they wanted \textit{``a map-based progress''} to track their journey through meditation (P004). Future systems should therefore move towards more innovative and tangible ways of tracking progress, such as incorporating visual timelines or dashboards that illustrate users' evolving goals, learned techniques, and engagement patterns.

\section{Discussion}
\label{sec:discussion}

In this work, we presented \projectname, a multi-agent system designed to deliver expert-aligned and personalized mindfulness meditation guidance. Our findings from {a formative and a four-week deployment study} demonstrate that \projectname significantly improved in-session engagement and self-awareness, reduced momentary stress, and, most critically, achieved a higher long-term engagement and greater improvements in mindfulness compared to a non-personalized baseline. 
In this section, we discuss the broader implications of our work, including the system's role in fostering an emotional companionship with users, the function of reflection as a bridge between formal practice and daily life, and the design tension between ensuring expert-aligned safety and providing the generative novelty required for long-term engagement. We also summarize the existing limitations of our work and suggest directions for future research.

\subsection{From Instruction to Companionship: Designing for Continuous, Empathetic Interactivity}
Our findings suggest that interactive personalization in \projectname did more than sustain engagement; it altered the user's relationship with the system, transforming it from a digital tool into an emotional companion. The system fostered this connection by providing space for users to express themselves and by creating a sense of \textit{``being known''}, essential for engagement according to mindfulness experts. It achieved this by recalling prior inputs, tailoring content to users' momentary states, and addressing them directly. Participants consistently noted that this contextual awareness made the practice feel more relevant and supportive, shifting \projectname's role from one of technical adaptation to a \textit{``companion''} capable of emotional support.

However, to fully realize the potential of AI as a long-term behavioral health partner, future systems must evolve beyond simple text-based recall toward \textit{socially intelligent and multi-modal companionship}. First, a robust companion must possess \textit{cultural and social awareness}. {While generic empathy and recall was appreciated, engagement diminished when the system failed to grasp social or cultural nuances, for example, when personalization felt forced or repetitive or when linguistic style did not align with users’ cultural backgrounds} \cite{liuUnderstandingPublicPerceptions2024, monteiroConversationalAgentsSurvey2023, bickmoreEstablishingMaintainingLongterm2005}. Future iterations should advance to understand the social fabric of the user's life, adapting its tone and advice based on cultural context, workplace dynamics, or social roles. A socially intelligent agent would not only validate emotions but also frame mindfulness interventions in a way that aligns with the user's specific values and reality.

Second, true companionship implies an intuitive understanding that does not always require explicit explanation. {Currently, users must actively journal to inform the system, creating a critical tension highlighted in our qualitative findings: although personalized outputs were valued, the required manual input was frequently experienced as onerous and time-intensive.} Future designs should integrate \textit{passive sensing and multi-modal data streams} to create a ``just-in-time'' understanding of the user's state that minimizes friction \cite{nahum-shaniJustinTimeAdaptiveInterventions2018a, hanStressBalPersonalizedJustintime2023}. By synthesizing physiological signals or contextual data, the system could infer stress and proactively offer support without requiring the user to articulate their distress. This shifts the interaction from a reactive tool to a proactive presence that understands the user's needs with minimal effort.

Finally, a companion relationship is dynamic, not static. Just as a relationship with a human teacher evolves from instruction to mentorship, the AI's role should undergo \textit{longitudinal evolution}. {Our qualitative findings indicate that users’ needs shift over time: early stages benefit from structured guidance, whereas later stages favor reduced friction and greater autonomy.} While our current system tracks history and allows adjustment of guidance level, future systems should model the trajectory of the relationship and growth. As the user progresses from a novice to an advanced practitioner, the AI's persona \cite{tsengTwoTalesPersona2024, gomezHowLargeLanguage2024} should shift dynamically—perhaps fading from a talkative instructor to a quiet, supportive peer. This evolution mirrors the depth of human connection, where shared history allows for more efficient communication and deeper emotional resonance over time.

\subsection{The Reflective Bridge: Grounding Meditation in Daily Experience} 
Our findings suggest that reflection is not merely an add-on feature, but the critical bridge that connects meditation with everyday mindful living. Participants frequently likened the reflective chatbot to a journaling exercise that helped them clarify feelings and reframe experiences, offering contextual cues that enabled them to remain present (Sec. \ref{results:qualitative_feedback}). This resonates with prior research showing that journaling practices, particularly those fostering attachment security and self-compassion, can enhance self-awareness and increase the likelihood of sustained engagement in meditation \cite{roweAttachmentSecuritySelfcompassion2016a}. By situating our findings within this theoretical grounding, we show how reflection may act as a scaffold for self-awareness, integrating deeply with participants' daily lives.

Future systems can extend this connection by freeing reflection from its current position as a pre-session initialization. Participants expressed a desire for reflection to be woven dynamically throughout the experience, such as potentially appearing in the middle of a mindfulness session as interactive check-ins, or at the end to consolidate insights. By expanding reflection to bookend the practice, systems can help users explicitly link their post-meditation state back to their initial intentions. Furthermore, to fully support a ``Narrative of Becoming,'' \cite{oysermanIdentitybasedMotivationImplications2009} reflection should evolve beyond text-based chats into more visual and concrete forms. {Our qualitative findings suggest a desire for representations that make progress tangible, such as visual summaries or structured overviews that illustrate how individual sessions connect to longer-term goals and lived experiences.} By visualizing these abstract mental shifts, future systems can make the intangible benefits of mindfulness concrete, reinforcing the user's motivation by showing them not just that they meditated, but how they are changing.

\subsection{A Design Paradox of Balancing Generative-LLM Safety and Diversity}
Moreover, our findings highlight a design paradox in using generative LLM for mental health intervention:  the need to provide diverse content for long-term engagement while operating within safe boundaries. To guarantee safety and efficacy, our system was built upon a foundation of expert-aligned, evidence-based safety templates. 
This structure is essential for ensuring the system's integrity and reducing the risk of harmful or inaccurate LLM-generated content. However, our findings also demonstrate that overly rigid structures may lead to repetition that potentially diminishes engagement over time.
As our qualitative feedback showed, users require novelty and variability to remain motivated, and they begin to disengage the moment content feels repetitive.

{To resolve this tension, future work should move beyond a binary choice between safe scripts and open generation. Instead, a \textit{Modular Content Architecture} that smartly compartmentalizes the intervention into static and dynamic components can be a potential solution \cite{dongBuildingGuardrailsLarge2024, dongSafeguardingLargeLanguage2024}.
In this architecture, the system distinguishes between \textit{clinical anchors} (core therapeutic mechanisms such as specific breathing ratios or cognitive re-framing techniques), which remain fixed to ensure safety, and \textit{narrative wrappers} (e.g., introductions, metaphors, and personalized homework), which are generative to optimize engagement. Furthermore, decomposing safety templates would allow designers to tune the fixed–versus–personalized balance of each component across domains and contexts. Rather than relying on monolithic scripts, future systems should break these templates down into granular sub-modules guided by human experts. For instance, an instruction for a body scan could be segmented into smaller, interchangeable micro-scripts that vary in tone and phrasing while adhering to the same clinical principle. This granular approach allows the LLM to assemble a fluid, highly personalized guidance flow that feels spontaneously generated, yet remains bounded by expert-defined safety guardrails. By dissecting the intervention in this way, designers can achieve ``principled improvisation'' \cite{wanniarachchi_personalization_2025}, creating an experience that is endlessly novel in its delivery but consistent in its therapeutic core.}

{\subsection{Future of Personalization: From Experience to Identity}
Our findings suggest that a promising potential of AI personalization lies in bridging the gap between momentary experience and long-term identity formation. Behavioral change is most durable when a user shifts from \textit{doing} a task to \textit{being} a practitioner \cite{oysermanIdentitybasedMotivationImplications2009}. In this work, we established a technical foundation for this shift through a multi-agent synergy: the Reflection Agent empowered users to co-create their guidance rather than passively consuming it; the Personalization Agent wove these insights into the session to resonate with the user's immediate identity; and the Expert-Alignment Agent ensured this exploration remained within a safe ``container'' of clinical efficacy.}

{Moreover, although \projectname is instantiated with the UM curriculum, its underlying mechanisms are curriculum-agnostic: framework-specific materials can be swapped for alternative evidence-based mindfulness protocols. Technically, this means that training materials, such as the Concept Introduction, Meditation Scripts, Instructor-Practitioner interactions, and expert-based feedback used to finetune the agents, can be updated to reflect other frameworks, while the underlying system remains unchanged. In this sense, UM serves as a concrete testbed for a more general pattern of expert-guided, identity-sensitive personalization. Future work should build upon this by explicitly designing for the psychological needs of autonomy, competence, and relatedness outlined in Self-Determination Theory \cite{ryanSelfDeterminationTheoryFacilitation}. By integrating the \textit{modular content architecture}, \textit{passive-sensing integration}, and \textit{socially intelligent memory} discussed earlier, future systems can evolve to mirror the user's growth back to them, transforming the AI from a simple content generator into a partner in personal evolution, where a series of isolated sessions coalesces into a coherent, user-authored story of becoming.}

\subsection{Limitations}
\label{sub:discussion:limitation}

Though effective, our study has several limitations. First, the two-week free practice phase, while providing initial engagement insights, was too brief to measure the sustained psychological benefits that typically emerge over a longer period of time. Future longitudinal studies are needed to assess these long-term outcomes. Besides, a longer deployment would enable the use of techniques like Reinforcement Learning from Human Feedback (RLHF) to continuously refine the personalization models, increasing the system's effectiveness over time.
Second, the time required for LLM processing introduced significant latency. While we controlled for this in our study by simulating comparable wait times across all conditions, these delays remain a practical barrier to a fluid user experience and could hinder real-world adoption.
Third, while the meditation content was personalized, the audio delivery was not, and the quality could be improved. Future iterations could aim to expand personalization to vocal styles and other prosodic elements like pacing and tone. This would allow the system to better match the delivery to a user's preferences and the emotional context of the session, creating a more seamless and deeply adaptive experience.
{Fourth, the four-week deployment study focused on a two-arm comparison (StaticAgent vs. \projectname) to prioritize the primary contrast, limiting our ability to assess intermediate variants. Future larger-scale deployments could systematically vary and ablate system components to more precisely characterize their individual and combined contributions.}

\section{Conclusion}
\label{sec:conclusion}
Sustained engagement with mindfulness meditation remains a significant barrier to its long-term benefits. In this work, we address this challenge with \projectname, a multi-agent system that delivers expert-aligned, personalized mindfulness guidance. Informed by three core design goals derived from collaboration with four mindfulness experts, \projectname integrates an (1) Expert-Alignment Agent to ensure safety and quality, (2) a Reflection Agent to foster self-awareness, and (3) a Personalization Agent to generate resonant, context-aware meditation scripts. 
Our evaluation demonstrates the efficacy of this approach. A lab study (N=13) revealed that \projectname was strongly preferred over a static baseline, significantly increasing in-session engagement, deepening awareness, while reducing stress. 
A subsequent 4-week in-the-wild study (N=62) showed that \projectname led to higher long-term engagement, mood, and level of mindfulness. Qualitative feedback highlighted the system's key benefits: the relevance of personalized guidance, the value of interactive companionship, and the diversity of content.
These results underscore the potential of LLM-powered systems to make mindfulness more personalized and engaging, offering a step toward scalable tools that support sustained mindfulness meditation practices.

\begin{acks}
Research reported in this publication was supported in part by the National Institute of Mental Health of the National Institutes of Health under Award Number R44MH134709 and K01MH123505, the National Institute of Diabetes and Digestive and Kidney Diseases under Award Number R01DK128114, the National Cancer Institute under Award Number R01CA236860, and the National Institute on Drug Abuse under Award Number P50DA054039. The content is solely the responsibility of the authors and does not necessarily represent the official views of the National Institutes of Health. We also thank the Columbia University Research Stabilization Grant for supporting this research project. We thank all participants for taking part in our study and the anonymous reviewers for their constructive feedback.
\end{acks}

\section*{Acknowledgment of AI Use}
    We acknowledge the use of Generative AI tools to improve the grammar, style, and readability of this manuscript. These tools were used in our system (see details in Sec. \ref{sec:system}) and text editing, but played no role in the data analysis, interpretation, or generation of the core findings presented.

\bibliographystyle{ACM-Reference-Format}
\bibliography{mindfulagents_refs_consolidated}

\appendix
\onecolumn

\section*{Appendix}
\label{appendix:}

\begin{figure*}[htb!]
  \centering
  \includegraphics[width=\linewidth]{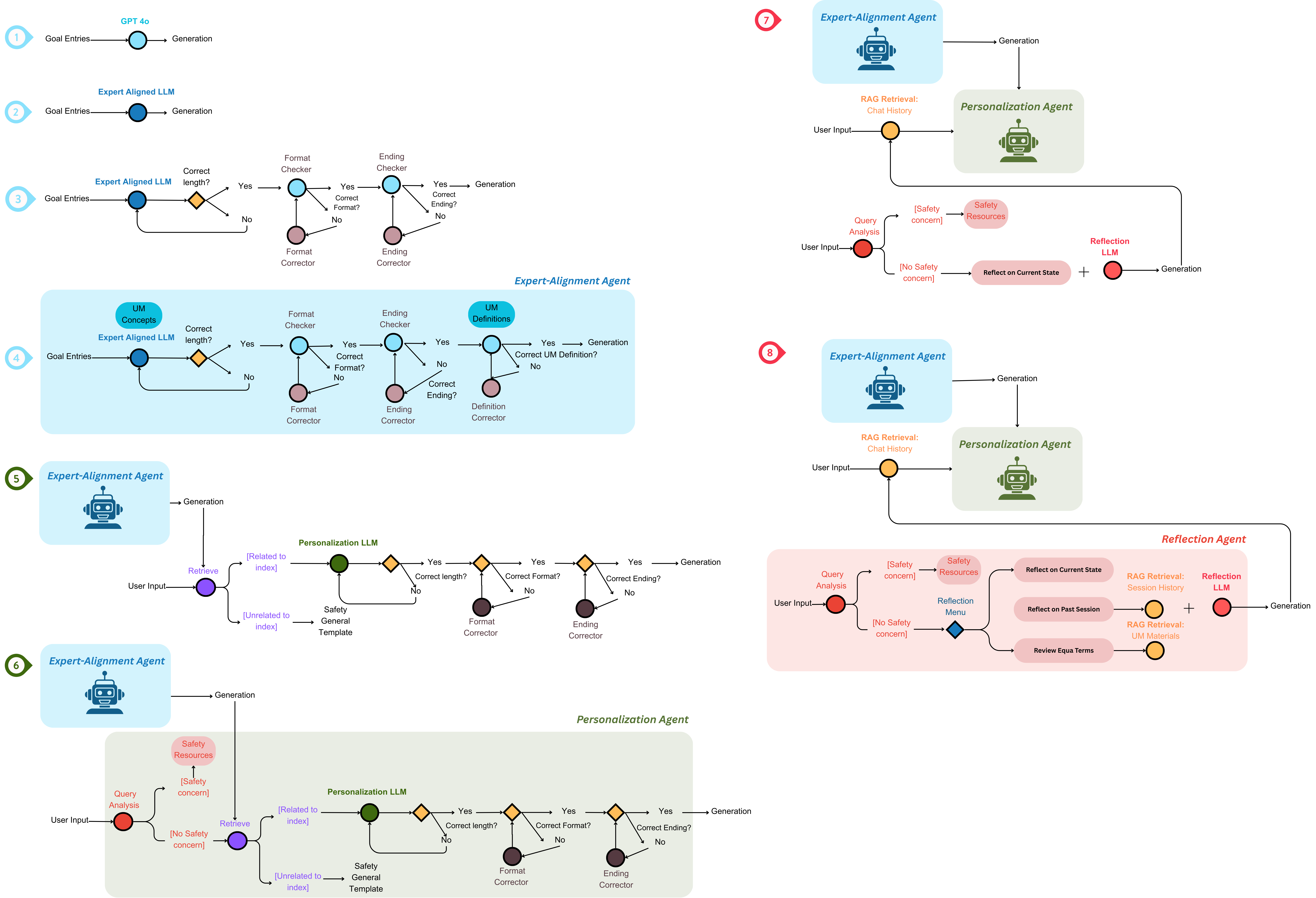}
  \caption{Overview of the iterative system design. Each panel shows one of eight design iterations, from a vanilla GPT\textendash4o baseline to the final \projectname architecture integrating Expert-alignment, Personalization, and Reflection agents.}
  \Description{A multi-panel diagram showing eight successive system iterations. Early panels depict a single LLM meditation coach with unstable outputs. Middle panels add finetuning and format checks to reduce hallucinations and enforce script structure. Later panels introduce separate agents for expert-aligned template creation, real-time personalization, and user reflection. The final panel shows the complete MindfulAgents pipeline combining expert templates, reflective chat, and personalized meditation generation.}
  \label{fig:design_iteration}
\end{figure*}

\begin{table*}[h]
\centering
\caption{Overview of design iterations leading to the final \projectname system.}
\label{tab:design-iterations}
\begin{tabularx}{\linewidth}{
  >{\raggedright\arraybackslash}p{0.07\linewidth}  %
  >{\raggedright\arraybackslash}p{0.22\linewidth}  %
  >{\raggedright\arraybackslash}X                  %
}
\toprule
\textbf{Iteration} & \textbf{Configuration} & \textbf{Key observations} \\
\midrule
1 & Vanilla GPT-4o
  & Vanilla GPT-4o prompted as a meditation coach. Extremely unstable: \textbf{frequent hallucinations, wrong format for the audio API, and inconsistent adherence to the Unified Mindfulness curriculum or expert guidance.} \\

2 & Expert-aligned finetuned model
  & GPT-4o finetuned on expert-written UM scripts. Reduced some obvious errors, but the model still \textbf{hallucinated and occasionally produced responses in an incorrect format for the audio API}. \\

3 & Finetuning + format checks
  & Added evaluation layers (script length, script format, and script ending) to validate and correct outputs before passing them to the audio API. Reliability improved substantially, but the \textbf{model still mishandled fine-grained details of the UM curriculum}. \\

4 & Expert-alignment agent
  & Restructured the system as an expert-aligned agent with additional curriculum alignment in evaluation and prompting. Eliminated most formatting and curriculum-consistency issues, but the multi-step agent pipeline introduced \textbf{high latency and increased risk of failure during long generations}, motivating extraction and caching of expert-aligned scripts. \\

5 & Expert-alignment agent + personalization LLM
  & Introduced a dedicated personalization LLM to adapt expert-aligned scripts to user-specific factors (e.g., core personalization inputs identified in DG~3) using faster deterministic checks. Improved responsiveness and personalization, but revealed the need for \textbf{stronger safety resources and clearer escalation paths for sensitive disclosures}. \\

6 & Expert-alignment agent + personalization agent
  & Expanded the personalization component into a full personalization agent with appropriate safety guardrails. Expert feedback highlighted the need for \textbf{richer user context and a stronger sense of therapeutic alliance}. \\

7 & Expert-alignment agent + personalization agent + initial reflection agent
  & Added an initial reflection agent to support post-session processing. This version primarily summarized the current session and did not yet provide rich, \textbf{in-person–like dialogue grounded in prior sessions or user questions}. \\

8 & \projectname 
  & Integrated an enhanced reflection agent that leverages current state check-in, past session reflection, and curriculum-related Q\&A. The final system delivers low-latency, curriculum-faithful, and personalized meditation guidance while supporting deeper reflective interaction. \\
\bottomrule
\end{tabularx}
\end{table*}

\begin{table*}
\centering
\renewcommand{\arraystretch}{1.5}
\caption{List of Unified Mindfulness (UM) techniques.}
\label{tab:um_techniques}
\resizebox{\textwidth}{!}{%
\begin{tabular}{p{2.5cm} p{12cm}}
\toprule
\textbf{UM Technique} & \textbf{Description} \\
\midrule
See-Hear-Feel       & A systematic way Unified Mindfulness organizes sensory experience into three categories: sight, sound, and body sensation. “See” includes outer sights of the world and inner visual thinking on the mental screen. “Hear” includes outer sounds of the world and inner auditory thinking, such as self-talk. “Feel” includes outer physical body sensations and inner emotional body sensations.  \\
\midrule
Restful States      & Any sensory experience characterized by stillness, quiet, or relaxation, either arising naturally or intentionally created. Includes visual rest such as light, darkness, or absence of images, auditory rest such as silence or a quiet mind, and physical or emotional rest such as relaxation, neutrality, or peace. Restful states support tranquility, help induce equanimity, and create a positive feedback loop that sustains motivation to practice.  \\
\midrule
Gone                & The instant you become aware that an experience is partially or completely ending, whether you’re clear about what the experience is or not.  \\
\midrule
Concentration Power & One of the three core attention skills, the ability to focus on what you choose when you want to. You can think of it as a kind of flexible focus.
Can be developed by focusing on a small or large object of focus, and for a short or long period of time. Can be developed by keeping your focus within a focus range for as long as possible (sustained concentration), or by focusing for brief moments on one or more sensory experiences within a focus range (momentary concentration).  \\
\midrule
Sensory Clarity     & One of the three core attention skills, the ability to track and explore sensory experience in real time. Has two sides: Detection (sensitivity to details) and Discernment (the ability to tell this from that, such as the distinction between See and Hear and Feel).
  \\
\midrule
Equanimity          & One of the three core attention skills, the ability to allow sensory experience to come and go without push and pull.
Can be developed through a range of methods, such as noticing when it happens spontaneously, dropping judgments, intentionally relaxing, non-interference, anchoring attention away, noticing disappearances, creating or collapsing distance between yourself and a focus object (through zooming, being the witness, etc.), or accessing a positive emotional state, such as gratitude, compassion, etc.  \\
\midrule
Noting & A widely used approach for developing CC\&E that involves acknowledging and focusing on sensory experience for brief moments of time which can last anywhere from a fraction of a second to thirty seconds or more (using what is known as momentary concentration). Labeling is an option to support the process. \\
\midrule
Labeling   & A technique option used to support noting; it typically involves thinking or saying aloud a word or phrase to describe what you’re noting; aids in the continuous application of CC\&E.  \\
\bottomrule
\end{tabular}
}
\end{table*}

\begin{table*}
\centering
\renewcommand{\arraystretch}{1.5}
\caption{List of category and goal pairs for goal-oriented practice scripts.}
\label{tab:category_goals}
\begin{tabular}{p{2.5cm} p{11cm}}
\toprule
\textbf{Category} & \textbf{Goal} \\
\midrule
Starting Day & Positivity, Express Yourself, Better Habits, Find Your Purpose, Baseline Happiness \\
\midrule
Ready to Work & Improve Focus, Peak Performance, Be More Present, Creative Work \\
\midrule
Taking a Break & Work Break, Rejuvenation, Gain Clarity \\
\midrule
SOS & Panic Attack, Pain, Challenging Situations \\
\midrule
Socializing & Support Others, Emotional Intimacy, Manage Conflict, Enjoy Social Life, Forgiveness \\
\midrule
Big Event & Public Speaking, Test / Interview, Difficult Conversation \\
\midrule
Ending the Day & Sleep, Deep Relaxation \\
\midrule
General & Surprise Me (\ie the system will randomly pick a mindfulness goal) \\
\bottomrule
\end{tabular}
\end{table*}

\begin{table*}
\footnotesize
\centering
\renewcommand{\arraystretch}{1.5}
\caption{List of ablation study persona and their profile information}
\label{tab:user_persona}
\resizebox{\textwidth}{!}{%
\begin{tabularx}{\textwidth}{l l X p{4cm}}
\toprule
\textbf{ID} & \textbf{Name} & \textbf{Persona Description} & \textbf{Goal} \\
\midrule
1 & Sarah & Sarah is a 34-year-old marketing director at a Fortune 500 company in Chicago. She works 60+ hour weeks, manages a team of 15 people, and frequently travels for business. Recently divorced with joint custody of her 8-year-old daughter, Sarah struggles to balance her demanding career with single motherhood. She experiences chronic stress, has trouble sleeping, and often feels like she's constantly running on empty. & Panic Attack (Study 1), Work Break (Study 2), Forgiveness and Letting Go (Study 3), Sleep (Study 4) \\
\midrule
2 & Marcus & Marcus is a 20-year-old pre-med student at a competitive university. He's the first in his family to attend college and feels immense pressure to succeed academically while working part-time to help support his family financially. Marcus experiences significant social anxiety, especially in group settings and during presentations. He has panic attacks during exams and often feels imposter syndrome among his peers. Having grown up in a traditional Latino household, he's somewhat skeptical about meditation but is desperate for relief from his constant worry and self-doubt. Marcus wants to find ways to calm his mind before tests, build confidence in social situations, and develop better emotional regulation skills.
 & Test / Interview (Study 1), Enjoy Social Life (Study 2), Improve Focus (Study 3), Test / Interview (Study 4) \\
\midrule
3 & Ellie & Ellie is a 67-year-old retired art teacher who has been exploring various spiritual practices since her husband passed away three years ago. She lives alone in a small coastal town in Maine and has plenty of time for contemplative practices. Ellie has tried yoga, attended meditation retreats, and reads extensively about mindfulness and Eastern philosophy. However, she often feels like she's "doing it wrong" and gets frustrated when her mind wanders during meditation. She's looking for deeper spiritual connection and meaning in her life, wanting to move beyond surface-level relaxation techniques.
 & Find Your Purpose (Study 1), Surprise Me (Study 2), Create More Positive Situations (Study 3), Forgiveness and Letting Go (Study 4) \\
\midrule
4 & David & David is a 28-year-old personal trainer and former Division I soccer player who lives in Austin, Texas. He's always been focused on physical performance and views his body as a machine to be optimized. Recently, David has been dealing with chronic pain from old sports injuries and has been recommended meditation by his physical therapist as part of his recovery. He's naturally skeptical of anything that seems "touchy-feely" or unscientific, preferring evidence-based approaches. David is also struggling with anger management issues that have affected his relationships and coaching style with clients.
 & Pain (Study 1), Peak Performance (Study 2), Support Others (Study 3), Improve Focus (Study 4) \\
 \midrule
 5 & Priya & Priya is a 45-year-old nurse practitioner who works in a busy emergency department in Phoenix. She's been on the frontlines of healthcare for over 15 years and has experienced significant burnout, especially after the pandemic. Priya is also the primary caregiver for her elderly mother who has dementia, while raising two teenage children with her husband. She feels emotionally drained, has developed compassion fatigue, and struggles with guilt about not being able to give her best to her patients, family, or herself. Priya has some familiarity with meditation through her Hindu cultural background but hasn't practiced regularly since childhood.
 & Support Others (Study 1), Deep Relaxation (Study 2), Gain Psychological Clarity (Study 3), Rebound from Challenging Situations (Study 4)  \\
\midrule
6 & Jimmy & Jimmy is a 31-year-old graphic designer and freelance artist living in Brooklyn, New York. He quit his corporate design job two years ago to pursue his passion for illustration and has been struggling financially ever since. Jimmy works irregular hours, often pulling all-nighters to meet client deadlines, and his income is unpredictable. He battles creative blocks, imposter syndrome, and the constant pressure to find new clients while maintaining his artistic integrity. Jimmy has experimented with recreational substances to enhance creativity but realizes this isn't sustainable. He's naturally introspective and has tried meditation apps sporadically but finds it hard to sit still when his mind is racing with creative ideas or financial worries.
 & Creative Work (Study 1), Express Yourself (Study 2), Deep Relaxation (Study 3), Rejuvenation (Study 4) \\
\bottomrule
\end{tabularx}
}
\end{table*}

\begin{table*}
\addtocounter{table}{-1}
\footnotesize
\centering
\renewcommand{\arraystretch}{1.5}
\caption{List of ablation study persona and their profile information (cont.)}
\label{}
\resizebox{\textwidth}{!}{%
\begin{tabularx}{\textwidth}{l l X p{4cm}}
\toprule
\textbf{ID} & \textbf{Name} & \textbf{Persona Description} & \textbf{Goal} \\
\midrule
7 & Becca & Becca is a 52-year-old administrative assistant from suburban Minneapolis whose youngest child just left for college. After 25 years of marriage, she and her husband are getting divorced, and she's living alone for the first time in her adult life. Becca has always defined herself through her roles as wife and mother, and now feels lost and unsure of her identity. She's dealing with perimenopause symptoms, including mood swings and sleep disturbances, and has gained weight that affects her self-esteem. Becca has never meditated before but has been recommended mindfulness by her therapist to help with the life transition. 
 & Find Your Purpose (Study 1), Forgiveness (Study 2), Rejuvenation (Study 3), Everyday Happiness (Study 4) \\
\midrule
8 & Michael & Michael is a 38-year-old associate professor of economics at a prestigious university who is up for tenure review next year. He's published extensively but constantly worries that his research isn't groundbreaking enough. Michael works 70+ hour weeks, has chronic insomnia, and experiences frequent tension headaches from stress. He's highly analytical and approaches everything, including his personal life, with rigorous logical thinking. Michael has read about meditation's cognitive benefits in scientific journals but has never tried it because he considers it "unscientific" and worries about losing his analytical edge. He's recently been having panic attacks before important presentations and realizes he needs help.
 & Public Speaking (Study 1), Improve Focus (Study 2), Express Yourself (Study 3), Public Speaking (Study 4) \\
 \midrule
 9 & Carmen & Carmen is a 24-year-old lifestyle influencer from Los Angeles with 200K followers across platforms. She built her brand around wellness and positivity but privately struggles with anxiety, depression, and the pressure to constantly perform happiness online. Carmen experiences severe FOMO, compares herself constantly to other influencers, and has developed an unhealthy relationship with social media metrics. She's tried various wellness trends for content but has never committed to a consistent meditation practice. Carmen feels like a fraud promoting wellness while struggling internally and fears that being authentic about her mental health struggles would damage her brand.
 & Gain Clarity (Study 1), Baseline Happiness (Study 2), Experience Closeness (Study 3), Enjoy Social Life (Study 4) \\
\midrule
10 & Bob & Bob is a 66-year-old former construction foreman from rural Montana who retired six months ago after 40 years in the trades. He's struggling with the transition from a physically demanding, structured work life to having unlimited free time. Bob feels useless and bored, misses the camaraderie of his work crew, and is experiencing mild depression. His wife suggests he needs new hobbies, but Bob feels too old to learn new things and worries about his declining physical abilities. He's skeptical of anything that seems "new age" or feminine, viewing meditation as something for "yoga people," not tough guys like himself. Bob has chronic lower back pain from years of physical labor and occasional anger outbursts that worry his family.
 & Pain (Study 1), Better Habits (Study 2), Sleep (Study 3), Ease Pain and Discomfort (Study 4) \\
\midrule
11 & Aisha & Aisha is a 26-year-old PhD candidate in psychology at a competitive research university. She's in her fourth year of graduate school, working on her dissertation while teaching undergraduate courses and conducting research. Aisha feels overwhelmed by academic pressure, imposter syndrome, and the uncertain job market in academia. She's passionate about her research on trauma therapy but finds herself emotionally affected by the heavy subject matter. Aisha comes from a working-class family and is the first to pursue higher education, feeling pressure to succeed not just for herself but to prove that her family's sacrifices were worthwhile. She has some experience with mindfulness through her psychology training but has never developed a personal practice.
 & Support Others (Study 1), Rejuvenation (Study 2), Work Break (Study 3), Support Others (Study 4) \\
\midrule
12 & Tony & Tony is a 29-year-old security guard who works the overnight shift at a corporate office building in Seattle. He's been working nights for three years to earn extra money while his wife attends nursing school during the day. Tony struggles with maintaining a healthy sleep schedule, often feeling groggy and disconnected from his family and friends who are on normal daytime schedules. He feels isolated and lonely during his shifts, often mindlessly scrolling social media to pass time. Tony has gained weight from irregular eating patterns and lack of exercise, and he's become increasingly irritable with his wife and young daughter during the few hours they're all awake together. He's never tried meditation but is desperate for something to help him feel more balanced.
 & Sleep (Study 1), Sleep (Study 2), Public Speaking (Study 3), Express Yourself (Study 4) \\
\bottomrule
\end{tabularx}
}
\end{table*}

\begin{table*}
    \centering
    \caption{Single-item measures used in the Formative Lab Study (Study~1). All items were rated on a 5-point Likert scale (1 = strongly disagree, 5 = strongly agree).}
    \label{tab:formative_measures}
    \begin{tabularx}{\linewidth}{lXX}
        \toprule
        \textbf{Metric} & \textbf{Questionnaire item} & \textbf{Explanation} \\
        \midrule
        In-session engagement &
        ``The session felt engaging.'' &
        Assesses how immersed, attentive, and focused participants felt during the meditation session. \\
        
        Perceived level of personalization &
        ``It addressed my current mood/goals.'' &
        Captures the extent to which participants felt the guidance was tailored to their present emotional state and intentions. \\
        
        Deepening self-awareness &
        ``It helped deepen my awareness.'' &
        Indicates whether the session supported noticing and understanding inner experiences, thoughts, and bodily sensations. \\
        
        Willingness for future use &
        ``I would choose to use this again.'' &
        Reflects participants' intention to use this type of session in the future, as a proxy for perceived long-term value. \\
        
        Stress reduction &
        ``My stress level reduced.'' &
        Measures perceived short-term relief from stress or tension following the session. \\
        \bottomrule
    \end{tabularx}
\end{table*}

\end{document}